\pdfoutput=1
\documentclass{svjour3}
\newcommand{\IT}{{\sffamily DUO}}
\usepackage{times}
\usepackage{cite}
 \usepackage{hyperref}
\usepackage{blindtext, graphicx}
\usepackage{hyperref}
\usepackage{tabu}
\usepackage{colortbl}
\usepackage{tikz}
\usepackage{wrapfig}
\usepackage{microtype}

\smartqed
\newcommand{\bi}{\begin{itemize}}

\newcommand{\ei}{\end{itemize}}

\newcommand{\tbl}[1]{Table~\ref{tbl:#1}}

\newcommand{\fig}[1]{Figure~\ref{fig:#1}}

\newcommand{\tion}[1]{\S\ref{tion:#1}}
\usepackage{balance}
\definecolor{Gray}{rgb}{0.88,1,1}
\definecolor{Gray}{gray}{0.85}
\definecolor{lightgray}{gray}{0.8}
\newcommand{\RED}{\color{black}}

\newcommand{\BLACK}{\color{black}}

\usepackage{subfig} 

\usepackage[linesnumbered,ruled,vlined]{algorithm2e}
\SetKwProg{Fn}{Function}{}{}

\usepackage{amsmath}

\usepackage[framed]{ntheorem}
\usepackage{framed}
\usepackage{tikz}
\usetikzlibrary{shadows}
\theoremclass{Lesson}
\theoremstyle{break}

\tikzstyle{thmbox} = [rectangle, rounded corners, draw=black,
fill=Gray!20,  drop shadow={fill=black, opacity=1}]

\newshadedtheorem{lesson}{Finding}

\usepackage{color}
\usepackage{xspace}
\definecolor{ScarletRed}{rgb}{0.80,0.00,0.00}

\begin{document}

    \title{Better Software Analytics via ``DUO'':\\Data Mining Algorithms Using/Used-by Optimizers}


\author{Amritanshu Agrawal \and Tim Menzies \and \\Leandro L. Minku
\  \and  Markus Wagner \and  Zhe Yu\thanks{Authors listed alphabetically.}
}


\institute{ Amritanshu Agrawal, Tim Menzies,  Zhe Yu  \at
              Department of Computer Science, North Carolina State University, Raleigh, NC, USA.
              \email{zyu9@ncsu.edu}           
            \and
           Leandro L. Minku \at
              School of Computer Science, University of Birmingham, Edgbaston, Birmingham, UK. 
              \email{l.l.minku@cs.bham.ac.uk}
           \and
           Markus Wagner \at
              School of Computer Science, University of Adelaide, Adelaide, SA, Australia.
              \email{markus.wagner@adelaide.edu.au}
}


\maketitle

\begin{abstract}
This paper claims that a new field of empirical software engineering research and practice is emerging: 
data mining using/used-by optimizers for empirical studies, or {\IT}. 
For example, data miners can generate models
that are explored by optimizers. Also, optimizers can advise how to
best adjust the control parameters of a data miner.
This combined approach    acts like an agent leaning over the shoulder
of an analyst
that advises ``ask this question next''  or ``ignore that problem, it is not relevant to your goals''. Further, those agents can help us build ``better''
predictive models, where ``better'' can be either greater predictive
accuracy or faster modeling time (which, in turn, enables the exploration of a wider range of options).
We also caution that  the era of papers that just use data miners
is coming to an end. Results obtained from an unoptimized data miner can be quickly refuted, just by applying an optimizer to produce a different (and  better performing) model. 
Our conclusion, hence, is that for software analytics it is  possible, useful and necessary to combine data mining 
and optimization using {\IT}.


\end{abstract}

\newpage
\section{Introduction}
\label{sect: Introduction}
After {\em collecting  data} about software projects,
and before {\em making conclusions} about those projects, there is a
middle step in empirical software engineering where the data is {\em interpreted}. 
When the
data
is very large and/or is expressed in terms of some complex model of software projects, then interpretation is often accomplished, in part, via
 some automatic algorithm.
For example, 
an increasing number of  empirical  studies 
base their conclusions on   data mining algorithms (e.g. see~\cite{27menzies2013,menzim18r,bird2015art,menzies2013data,2016tim})
  or model-intensive algorithms such as optimizers (e.g. see
  the recent  section on Search-Based Software Engineering in the December 2016 issue of this journal~\cite{Kessentini16}).

This paper asserts that, for software analytics it is 
{\em possible}, {\em useful} and 
{\em necessary} to combine data mining and optimization.
We call this combination {\IT}, short for \underline{d}ata miners \underline{u}sing/used-by
\underline{o}ptimizers. 
Once data miners and optimizers are combined, this results in  a  very different and interesting class of interpretation methods for empirical SE data.
  {\IT}     acts like an agent leaning over the shoulder
of an analyst
that advises ``ask this question next''  or ``ignore that problem, it is not relevant to your goals''. Further, {\IT} helps us build ``better''
predictive models, where ``better'' can be for instance greater predictive
accuracy, or models that generalize better, or faster modeling time (which, in turn, enables the faster exploration of a wider range of options). Therefore, {\IT} can speed up and produce more reliable analyses in empirical studies.

\begin{wrapfigure}{r}{2.5in}
 \includegraphics[width=2.5in]{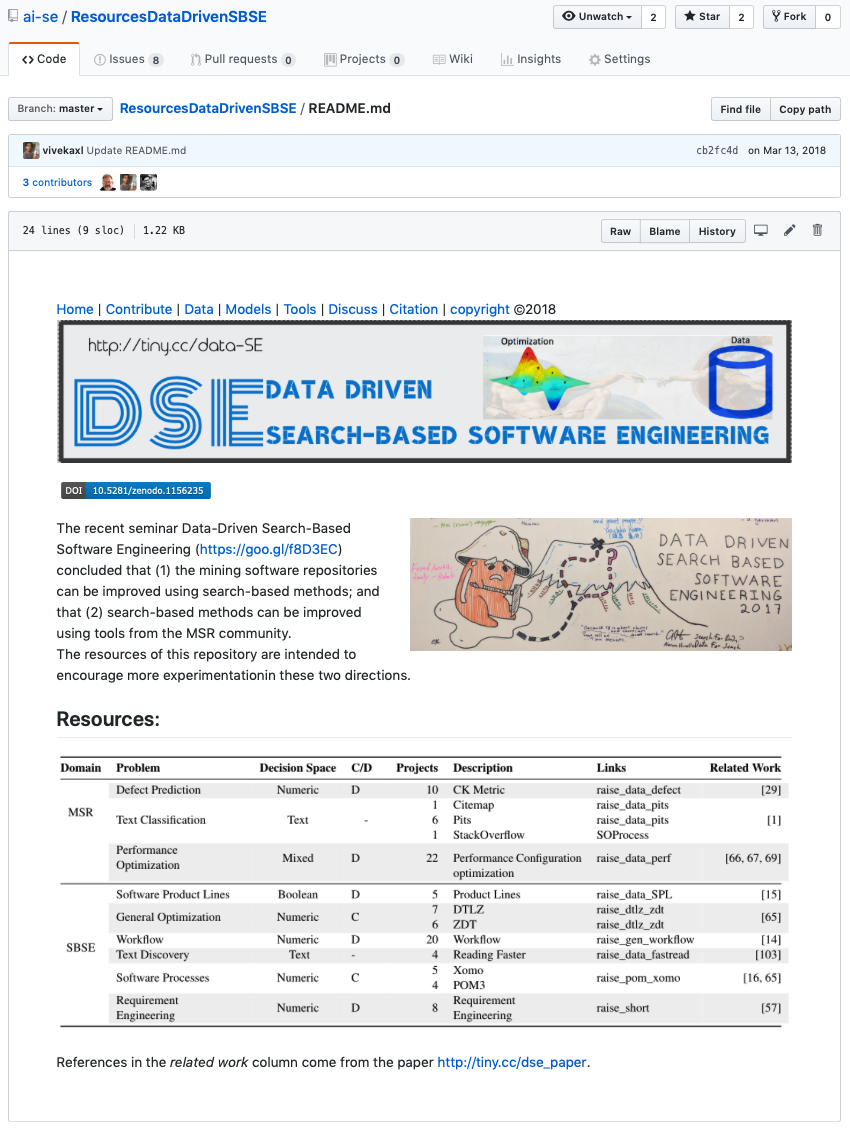}
\caption{  Training and teaching resources for this work: \href{http://tiny.cc/data-se}{http://tiny.cc/data-se}. The authors of this paper
invite the international research community to issue pull requests on this material. }\label{fig:train}
\end{wrapfigure}

This paper makes the following claims about {\IT}:  
 \begin{itemize}
\item {\bf Claim1:} {\em \RED For software engineering tasks,{\BLACK} optimization and data mining
are very similar}.  Hence, it is natural
and simple to combine the two methods.
\item {\bf Claim2:} {\em \RED For software engineering tasks.{\BLACK} optimizers can greatly improve data miners}. A data miner's default tuners can lead to sub-optimal performance. Automatic optimizers
can find tunings that dramatically improve that performance~\cite{35fu2016,54agrawal2018better,57agrawal2018wrong}. 
\item {\bf Claim3:}  {\em \RED For software engineering tasks,{\BLACK} data miners   can greatly improve optimization.}  If a data miner groups together related items, an optimizer 
can explore and report conclusions that are general across a set of solutions.
Further, optimization for SE problems can be very slow (e.g. consider the 15 
years of CPU needed by Wang et al.~\cite{wang2013searching}).
But if that optimization executes over the groupings found by a data miner,  that inference can terminate orders of magnitude faster ~\cite{90nair2018finding,krall2015gale}.
\item {\bf Claim4:} 
{\em \RED For software engineering tasks,{\BLACK} data mining without optimization is not recommended.} Conclusions reached from an unoptimized data miner can be  changed, sometimes even dramatically improved, 
by running the same tuned learner on the same data~\cite{35fu2016}.  
Researchers in data mining should, therefore, consider adding an optimization step
to their analysis. 

\end{itemize}
This  paper extends a prior conference paper
on the same topic~\cite{Nair:2018}. That prior paper   focused on case study material
for {\IT}-like applications (see \fig{train}). While a useful resource, it did not
place {\IT} in a broader context. Nor did it contain the   literature review of this paper. That is, this paper is both more general (discussing the broader context) and more specific (containing a detailed literature review) than prior work.

The rest of this paper is organized as follows. 
The next section describes some related work.
After that, we devote one section to each of   
{\bf Claim1, Claim2, Claim3} and 
{\bf Claim4}. To defend these claims we use evidence drawn from  a   literature review
of applications of {\IT}, described in Table~\ref{tbl:lt}
(\tbl{optimizers} and \tbl{learn} offer notes on the data miners and optimizers seen in the literature review). Finally, a {\em Research Directions} section discusses numerous
open research issues that could be better explored within the context of {\IT}.

 \RED
 \section{Tutorial}\label{tutorial}
 
 Before discussing combinations of data miners and optimizers,
we should start by defining each term separately, and discussing their relationship.
{\em Optimizers} reflect over a model to learn inputs that most
improve model output.
{\em Data miners}, on the other hand,
reflect over data to return a
summary of that data. 
Data miners usually explore a fixed
set of data while optimizers can generate more data by re-running the model $N$ times using different inputs.
Data miners ``slice''  data such that similar patterns  are found within each division.
Optimizers  ``zoom'' into interesting regions of the data,
using a model to  fill  in   missing details
about those regions. 
 
There are many different kinds of data miners, each of which might produce a different kind of model. For example, regression
methods generate equations;  nearest-neighbor-based methods 
might yield a   small set of most interesting
attributes and examples;
and  neural net methods return a weighted directed graph.
Some data miners generate combinations of models. For example,
the M5 ``model tree learner'' returns a set of equations
and a tree that decides when to use each equation~\cite{Quinlan92learningwith}.
  \begin{table}[p]
{\small
\begin{tabular}{|l|p{.3\linewidth}|p{.57\linewidth}|}\hline
Research &  \multicolumn{2}{p{.87\linewidth}|}{Q1. What papers have used DUO in the past?}\\
questions & \multicolumn{2}{p{.87\linewidth}|}{Q2. When were they published?}\\
 & \multicolumn{2}{p{.87\linewidth}|}{Q3. What problems from what software engineering domains have they tackled?}\\
  & \multicolumn{2}{p{.87\linewidth}|}{Q4. What optimizers have they used?}\\
 & \multicolumn{2}{p{.87\linewidth}|}{Q5. What data miners have they used?}\\
  & \multicolumn{2}{p{.87\linewidth}|}{Q6. What were the advantages offered by DUO?}\\ \hline
Search  & \multicolumn{2}{p{.87\linewidth}|}{software engineering AND (``optimization'' OR ``evolutionary algorithm'') AND  }\\
query   & \multicolumn{2}{p{.87\linewidth}|}{( ``data mining'' OR  ``analytics'' OR ``machine learning'')} \\ \hline
Search  & \multicolumn{2}{p{.87\linewidth}|}{Three widely used literature sources were adopted:}\\ 
engines & \multicolumn{2}{p{.87\linewidth}|}
{\vspace{-4mm}
\begin{itemize}
\item ACM Guide to Computing Literature (\url{https://dl.acm.org/advsearch.cfm?coll=DL&dl=ACM})
 \item IEEEXplore (\url{https://ieeexplore.ieee.org/search/advsearch.jsp})
 \item Google Scholar (\url{https://scholar.google.com/\#d=gs_asd&p=&u=})
\end{itemize}}\vspace{-3mm}\\ \hline

 Inclusion  & \multicolumn{2}{p{.87\linewidth}|}{\vspace{-3mm}\begin{itemize}\item  ACM and IEEEXplore: $>$3 citations per year OR published in 2017/2018.\end{itemize}}\vspace{-4mm}\\
criteria &    \multicolumn{2}{p{.87\linewidth}|}{\vspace{-3mm}\begin{itemize}
    \item  Google Scholar: ($>$10 cites/year OR published in 2017/2018) AND in first 20 result pages.
    \item For all search engines:  paper relates to software engineering  and must  use DUO.
    \end{itemize}}\vspace{-2mm}\\
  & \multicolumn{2}{p{.87\linewidth}|}{The number of citations was more strict in Google Scholar, because this search engine usually retrieves more citations than the others. We restricted the Google Scholar results to the first 20 pages (200 papers) because Google Scholar allows papers that do not match the search query completely to be retrieved, resulting in 28,000 results that would need to be manually filtered for the software engineering domain and number of citations per year.}\\ \hline
Q1 & \multicolumn{2}{p{.87\linewidth}|}{The search query retrieved 393 results when considering all ACM, IEEEXplore and the first 20 pages of Google Scholar results. After excluding papers that did not match the citation and year of publication criteria, we obtained 90 papers. After excluding any duplicates and papers that were not in the software engineering domain, we obtained 72 papers. Finally, among these 72, 29 used DUO. These are: \cite{1ali2018,2sadiq2018,13huang2006,14zhong2004,19kumar2008,20huang2008,26chiu2007,31chen2018,35fu2016,38minku2013,40sarro2016,42oliveira2010ga,43panichella2013effectively,44liu2010evolutionary,54agrawal2018better,55de2010symbolic,57agrawal2018wrong,60fu2017easy,64abdessalem18,87nair2017using,88majumder2018500+,89chen2018sampling,90nair2018finding,93du2015evolutionary,94minku2013analysis,97sarro2012further,98sarro2012single,100feather2002converging,102treude2018per}. In addition, we found a   literature review related to DUO \cite{30afzal2011}. This   review is on the topic of genetic programming for predictive modeling in software engineering. Genetic programming can be seen as an optimization algorithm.}\\ \hline
Q2 & \multicolumn{2}{p{.87\linewidth}|}{The number of DUO papers published per year is shown below, with 2018 having the largest number of papers. One might claim that this is because we ignore the number of citations in 2018. However, this is also the case for 2017, which had a considerably smaller number of DUO papers. Therefore, DUO seems to have been recently attracting increased research interest.}\\ 
 & 
\multicolumn{2}{p{.87\linewidth}|}{
\vspace{.4mm}{\scriptsize  
\hspace{3cm}\begin{tabular}{rrl}
2002 & 1 &\rule{3.3mm}{2mm}\\
2004 & 1 &\rule{3.3mm}{2mm}\\
2006 & 1 &\rule{3.3mm}{2mm}\\
2007 & 1 & \rule{3.3mm}{2mm}\\
2008 & 2 & \rule{6.6mm}{2mm}\\
2010 & 3 & \rule{9.9mm}{2mm}\\
2012 & 2 & \rule{6.6mm}{2mm}\\
2013 & 3 & \rule{9.9mm}{2mm}\\
2015 & 1 & \rule{3.3mm}{2mm}\\
2016 & 2 & \rule{6.6mm}{2mm}\\
2017 & 2 & \rule{6.6mm}{2mm}\\
2018 & 10 & \rule{33mm}{2mm}\\
\end{tabular}}\vspace{.5mm}}\\ \hline
Q3 & Domain where {\IT} is applied: & Specific problem: \\ 
 & \cellcolor{blue!10} Project management & \cellcolor{blue!10} Software effort estimation \cite{13huang2006,19kumar2008,20huang2008,26chiu2007,38minku2013,40sarro2016,42oliveira2010ga,94minku2013analysis,98sarro2012single,89chen2018sampling}, managing human resources \cite{31chen2018}. \\
 & Requirements &  Requirements optimization \cite{100feather2002converging}.\\
 & \cellcolor{blue!10} Design &  \cellcolor{blue!10}Software architecture optimization \cite{93du2015evolutionary}, extraction of products from very large product lines \cite{89chen2018sampling}.\\
 & Security &  Intrusion detection\cite{1ali2018,2sadiq2018}\\
 & \cellcolor{blue!10}Software quality & \cellcolor{blue!10}Software detect prediction \cite{14zhong2004,35fu2016,44liu2010evolutionary,54agrawal2018better,55de2010symbolic,97sarro2012further,47tantithamthavorn2016automated}, test case generation \cite{64abdessalem18}.\\
 & Software configuration &  Software configuration optimization \cite{90nair2018finding,87nair2017using,31chen2018}.\\
 & \cellcolor{blue!10}Text mining: StackOverflow &  \cellcolor{blue!10}Linking posts \cite{60fu2017easy,57agrawal2018wrong}, topic modeling \cite{57agrawal2018wrong,102treude2018per}.\\
 & Text mining: Defect Reports &  Defect reports topic modeling \cite{57agrawal2018wrong}.\\
 & \cellcolor{blue!10}Text mining: Software Artifact Search, Linking and Labeling & \cellcolor{blue!10} Traceability link recovery \cite{43panichella2013effectively}, locate features in source code \cite{43panichella2013effectively}, software artifacts labeling \cite{43panichella2013effectively}, topic modeling of software engineering papers \cite{57agrawal2018wrong}, Stack Overflow/GitHub topic modeling \cite{102treude2018per}. \\ \hline
Q4 & \multicolumn{2}{p{.87\linewidth}|}{See Table \ref{tbl:optimizers}.}\\ \hline
Q5 & \multicolumn{2}{p{.87\linewidth}|}{See Table \ref{tbl:learn}.}\\ \hline
Q6 & \multicolumn{2}{p{.87\linewidth}|}{See  Sections \ref{sec:claim3}, \ref{sec:claim4} and \ref{sec:claim5}.}\\\hline
\end{tabular}}
\caption{Exploring the DUO literature.}\label{tbl:lt}
\end{table}


\begin{table}[!b]
{\small
\begin{tabular}{|p{.99\linewidth}|}\hline
\rowcolor{blue!10}
\textbf{Genetic Algorithms (GAs)} execute over multiple  generations. Generation zero is usually initialized at random. After that, in each generation,
candidate items are  subject to {\em select} (prune away the less interesting solutions), 
{\em crossover} (build new items by combining parts of selected items), and {\em mutate} (randomly
perturb part of the new solutions). Modern GAs take different approaches to the {\em select} operator e.g. dominance rank, dominance count, and dominance depth. Notable exceptions are MOEA/D that use a decomposition operator to divide all the solutions into many small  neighborhoods where if anyone finds a better solution, all its neighbors move there as well~\cite{anderson2005practical,13huang2006,26chiu2007,40sarro2016,42oliveira2010ga,43panichella2013effectively,93du2015evolutionary,94minku2013analysis,97sarro2012further}. 
\\

\textbf{Different evolution (DE)} execute over multiple  generations. Generation zero is usually initialized at random. After that, in each generation,
candidate items are  subject to {\em select} (prune away the less interesting solutions), 
{\em mutate} (build new items by combining with 3 other random candidates from the same generation)~\cite{Storn:1997,35fu2016,54agrawal2018better,57agrawal2018wrong,60fu2017easy,88majumder2018500+}.
\\
\rowcolor{blue!10}
\textbf{MOEAs} contains different types of multi-objective evolutionary algorithms such as MOEA/D, NSGA-II, and more. They differ based on either the diversity mechanism (such as crowding distance or hypervolume contribution), or their sorting algorithm, or their use of target vectors, etc.
\cite{huang2005multiobjective,38minku2013,deb02,31chen2018,40sarro2016,64abdessalem18,Chand2015manyemo}.
\\

\textbf{Particle Swarm Optimization (PSO)} works by having a swarm of candidates where these candidates move around in the  search space using simple formulae. The movements of the particles are guided by their own best-known position in the search-space as well as the entire swarm's best-known position. When improved positions are being discovered these will then come to guide the movements of the swarm~\cite{poli2007particle,1ali2018,2sadiq2018,55de2010symbolic}.
\\
\rowcolor{blue!10}
\textbf{Genetic programs (GPs)} are like GAs except that while GAs manipulate candidates that are lists of options, GPs manipulate items that are trees. GPs can, therefore, be better for problems with some recursive structure (e.g. learning the parse tree of a useful equation) or when human-readable models are sought~\cite{banzhaf1998genetic,koza1994genetic,30afzal2011}.
\\

\textbf{SWAY} first randomly generates a large number of candidates, recursively divides the candidates and only selects one. SWAY quits after the initial generation while GA reasons over multiple generations. It makes no use of reproduction operators so there is no way for lessons learned to accumulate as it executes~\cite{31chen2018}.
\\
\rowcolor{blue!10}
\textbf{Tabu Search} uses a local or neighborhood search procedure to iteratively move from one potential solution x to an improved solution x' in the neighborhood of x, until some stopping criterion has been satisfied~\cite{glover1998tabu,31chen2018}.
\\
\textbf{FLASH}, a sequential model-based method such as Bayesian optimization, is a useful
strategy to find extremes of an unknown objective. FLASH is
efficient because of its ability to incorporate prior belief as
already measured solutions (or configurations), to help direct
further sampling. Here, the prior represents the already
known areas of the search (or performance optimization) problem. The prior can be used to estimate the rest of the
points (or unevaluated configurations). Once one (or many) points are evaluated based on the prior,
the posterior can be defined. The posterior captures the updated belief in
the objective function. This step is performed by using a
machine learning model, also called surrogate model ~\cite{90nair2018finding}. \\ \hline


\vspace{-1mm}
{\scriptsize  
\hspace{2.4cm}\begin{tabular}{rrl}
Genetic Algorithm & 31\% &\rule{31mm}{2mm}\\
MOEAs & 15\% & \rule{15mm}{2mm}\\
Differential Evolution & 20\% &\rule{20mm}{2mm}\\
Particle Swarm Optimization & 12\% & \rule{12mm}{2mm}\\
Genetic Programming & 7\% & \rule{7mm}{2mm}\\
Tabu Search & 4\% & \rule{4mm}{2mm}\\
SWAY & 7\% & \rule{7mm}{2mm}\\
Flash & 4\% & \rule{4mm}{2mm}\\
\end{tabular}
}\\\hline
\end{tabular}}
\caption{Notes on different optimizers. Bar chart at bottom shows the frequency of  use  in papers of  Table~\ref{tbl:lt}'s literature review.}\label{tbl:optimizers}
\end{table} 
  Whereas data miners
usually have ``hard-wired goals'' (e.g. improve accuracy),
the goals of an optimizer can be adjusted from problem 
to problem. In this way, an optimizer
can be tuned to the needs of 
different business users. For example:
\bi
\item
For requirements engineers, we can find the 
{\em least} cost
mitigations that enable the delivery of {\em   most} requirements~\cite{100feather2002converging}.
\item 
For project managers, we can apply optimizers
to software process models to find options that deliver {\em more} code in {\em less} time with {\em fewer bugs}~\cite{Menzies:2007}.
\item 
For  developers, our optimizers can tune
data miners looking for ways to find {\em more} bugs
in {\em fewer} lines of code (thereby reducing the 
human inspection effort required  once the learner has finished~\cite{Fu2016Tuning}.
\item
Etc.
\ei

In any engineering discipline, including software engineering, it is common to trade-off between multiple
completing goals (e.g.  all the   examples in the above  list are competing for multi-objective goals).
 Hence, for the rest of this tutorial section, we focus on multi-objective optimization.

  \subsection{Multi-Objective Optimization}

\begin{table}

{\small
 
\begin{tabular}{|p{.99\linewidth}|}\hline

\rowcolor{blue!10}
\textbf{Decision Tree learners} such as CART and C4.5 seek attributes which, if we split on their ranges,
most {\em reduces} the  expected value of the diversity in splits. These algorithms then recurse over each split to find further useful divisions. For numeric classes, diversity may be measured in terms of variance.
For discrete classes, the Gini or entropy measures can assess diversity. Decision tree learners are widely applied in software engineering due to the simplicity and interpretability~\cite{22catal2009,24vandecruys2008,26chiu2007,35fu2016,40sarro2016,42oliveira2010ga,44liu2010evolutionary,47tantithamthavorn2016automated,54agrawal2018better,55de2010symbolic,64abdessalem18,87nair2017using,90nair2018finding,94minku2013analysis}.
 \\
\textbf{Support Vector Machines} (SVMs) are supervised learning trying to separate training items from two classes by a clear gap~\cite{steinwart2008support}. For linearly non-separable problems, SVMs either allow but penalize misclassification of training items (soft-margin) or utilize kernel tricks to map input data to a higher-dimensional feature space before learning a hyperplane to separate the two classes. Many software engineering researchers have explored using SVM models to predict software artifacts~\cite{24vandecruys2008,42oliveira2010ga,47tantithamthavorn2016automated,54agrawal2018better,55de2010symbolic,60fu2017easy,88majumder2018500+,97sarro2012further}.
\\
\rowcolor{blue!10}
\textbf{Instance-based algorithms}: instead of fitting a model on the training data, instance-based algorithms stores all the training data as a database. When a new test item comes, the similarities between the test item and every training item are measured. The test item is then classified to the class where most of its similar training items belong to. Examples of instance-based algorithms include k-nearest neighbor~\cite{24vandecruys2008,40sarro2016,47tantithamthavorn2016automated,54agrawal2018better} and analogy algorithms~\cite{13huang2006,26chiu2007,94minku2013analysis,40sarro2016}.
\\
\textbf{Ensemble algorithms} use multiple learning algorithms to obtain better predictive performance than could be obtained from any of the constituent learning algorithms alone~\cite{polikar2006ensemble}. Usually, an ensemble algorithm builds multiple weak models that are independently trained and combines the output of each weak model in some way to make the overall prediction. Examples of ensemble algorithms include boosting~\cite{freund1996experiments,47tantithamthavorn2016automated}, bagging~\cite{breiman1996bagging,47tantithamthavorn2016automated,94minku2013analysis}, and random forest~\cite{liaw2002classification,22catal2009,35fu2016,54agrawal2018better,102treude2018per}.\\
  \rowcolor{blue!10}
\textbf{Bayesian algorithms} collect separate statistics for each class. Those statistics are used to estimate the prior distribution and the likelihood of each item in each class. When classifying a new test item, the estimated prior distribution and likelihood are applied to calculate its posterior distribution, which is then used to predict the class of the test item. Bayesian algorithms are widely applied in solving classification problems in software engineering~\cite{1ali2018,2sadiq2018,22catal2009,42oliveira2010ga,47tantithamthavorn2016automated,54agrawal2018better,55de2010symbolic}.
  \\
\textbf{Regression algorithms} use predefined functions to model the mapping from input space to output space. Parameters of the predefined function are estimated by minimizing the error between the ground truth outputs and the function outputs. Regression algorithms can be applied to solve both regression (e.g. linear regression~\cite{40sarro2016}) and classification problems (e.g. logistic regression~\cite{24vandecruys2008,35fu2016,47tantithamthavorn2016automated,54agrawal2018better,40sarro2016}).
\\
\rowcolor{blue!10}
\textbf{Artificial Neural Networks} (ANNs) are models that are inspired by the structure and function of biological neural networks~\cite{van2018artificial}. An ANN is based on a collection of connected units or nodes called artificial neurons, which loosely model the neurons in a biological brain. Each connection, like the synapses in a biological brain, can transmit a signal from one artificial neuron to another. An artificial neuron that receives a signal can process it and then signal additional artificial neurons connected to it. Such ANNs are usually applied in software engineering as baseline algorithms~\cite{26chiu2007,47tantithamthavorn2016automated,55de2010symbolic,94minku2013analysis}.
 \\
\textbf{Dimensionality reduction algorithms} transform the data in the high-dimensional space to a space of fewer dimensions~\cite{roweis2000nonlinear}. The resulting low-dimensional space can be used as features to train other data mining models or directly used as a clustering result. Examples of dimensionality reduction algorithms include principal component analysis~\cite{jolliffe2011principal}, linear discriminant analysis~\cite{riffenburgh1957linear,47tantithamthavorn2016automated}, and latent Dirichlet allocation~\cite{blei2003latent,43panichella2013effectively,57agrawal2018wrong,102treude2018per}.
 \\
  \rowcolor{blue!10}
\textbf{Covering (rule-based) algorithms} provide mechanisms that generate rules in a bottom-up, separate-and-conquer manner by concentrating on a specific class at a time and maximizing the probability of the desired classification. Examples of rule-based algorithms include PRISM~\cite{kwiatkowska2011prism} and RIPPER~\cite{cohen1995fast,24vandecruys2008,47tantithamthavorn2016automated,55de2010symbolic}.
  \\
\textbf{Deep Learning} methods are a modern update to ANNs that exploit abundant cheap computation. Deep learning uses a cascade of multiple layers of nonlinear processing units for feature extraction and transformation. Each successive layer uses the output from the previous layer as input. In a supervised or unsupervised manner, it learns multiple levels of representations that correspond to different levels of abstraction; the levels form a hierarchy of concepts~\cite{deng2014deep}. Examples of instance-based algorithms include deep belief networks and convolutional neural network~\cite{60fu2017easy}.
 \\
\hline
\vspace{1mm}
{\scriptsize  
\hspace{2.4cm}
\begin{tabular}{rrl}
Decision tree learners & 24\% &\rule{24mm}{2mm}\\
Support Vector Machines& 13\% &\rule{13mm}{2mm}\\
Instance-based algorithms & 11\% & \rule{11mm}{2mm}\\
Ensemble algorithms & 11\% & \rule{11mm}{2mm}\\
Bayesian algorithms & 11\% & \rule{11mm}{2mm}\\
Regression algorithms & 10\% & \rule{10mm}{2mm}\\
Artificial Neural Networks & 7\% & \rule{7mm}{2mm}\\
Dimensionality reduction algorithms& 7\% & \rule{7mm}{2mm}\\
Covering (rule-based) algorithms& 5\% & \rule{5mm}{2mm}\\
Deep Learning & 2\% & \rule{2mm}{2mm}\\
 \end{tabular}
}\\\hline
\end{tabular}}
\caption{Notes on the different data miners found by the literature review of \tbl{lt}. Bar chart at bottom shows the frequency of use in papers of Table~\ref{tbl:lt}'s literature review.}\label{tbl:learn}
\end{table}

All the optimizers of \tbl{optimizers} seek the maxima or minima of an objective functions {$y_i=f_i(x)$, $1 \leq i \leq m$
(where  $f_i$ are the objective 
 or ``evaluation'' functions, $m$ is the number of objectives,} $x$ is called the \textit{independent variable}, and $y_i$ are the \textit{dependent variables}).
                           
                               For single-objective ($m=1$) problems, algorithms aim at finding a single solution able to optimize the objective function, i.e., a \textbf{global maximum/minimum}.

                                  For multi-objective ($m>1$) problems,  
                                  there is no single `best' solution, but a number of `best' solutions.    
                                  These best solutions,
                                also known as the \textbf{Pareto Front}, are 
                                the {non-dominated solutions} found using a \textit{dominance} relation.      
                               This \textbf{domination criterion}  can be defined many ways:

                       \bi
                               \item The standard ``Boolean dominance'' predicates says that
                                 $x_1$ dominates 
                                another   $x_2$,
                                if $x_1$ is   better than $x_2$ in at least one objective and if $x_1$ is no worse than $x_2$ in all other objectives.
                                \item Another style of domination predicate is the ``Zitler indicator'' shown in Algorithm~\ref{cdom}. This method reports what loses least: moving from $x$ to $y$
or $y$ to $x$
(and the solution $x$ is preferred if moving to $x$ results
is the least  loss; see Algorithm~\ref{cdom}).
                                \item
                                Regardless of how domination is defined, a solution is called non-dominated if no other solution dominates it.
                                \ei

\begin{algorithm}[!t]
{\scriptsize
\begin{verbatim}
def x_better_than_y( 
            x,y,     # lists of candidate solutions
            weights, # dictionary of objective weights,
            goals,   # list of objective indexes
            lo, hi): # lists of low,high values of objectives  
    xloss, yloss, n = 0, 0, len(w)
    for g in goals:
        a      = normalize( x[g], lo[g], hi[g] ) 
        b      = normalize( y[g], lo[g], hi[g] )
        w      = weights[g]
        xloss -= 10**( w * (a-b)/n 
        yloss -= 10**( w * (b-a)/n )
    end
    return xloss < yloss

def normalize(z,lo,hi): return  (z  - lo) / (hi - lo + 0.00001)  
\end{verbatim}
}
\caption{Zitzler's indicator domination predicate~\cite{Zitzler2004IndicatorBasedSI}.
Most useful when reasoning about more than two objectives~\cite{33sayyad2013}.
In the weights array,  -1,1 means "minimize, maximize" respectively.
Objective scores are normalized 0..1 since, otherwise,
the exponential calculation might explode.}\label{cdom}
\end{algorithm}

        \begin{figure}[!b]
        \begin{center}  \includegraphics[height=4.8cm]{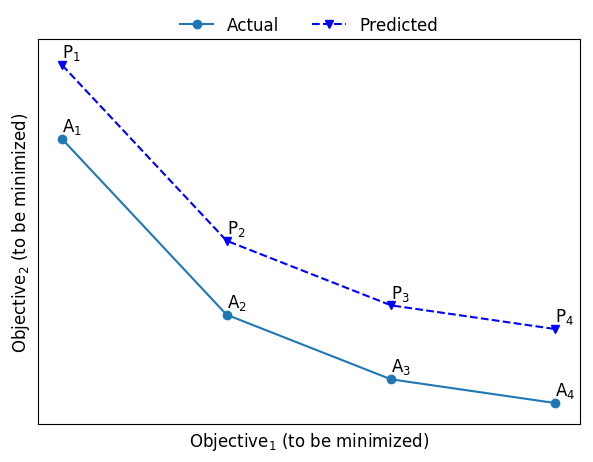}\end{center}
                \caption{Exploring multi-objective goals. Here, ``predicted'' is what is achieved and ``actual'' is some
                idealized set of goals (which may not be reachable). }\label{goals}                \end{figure}

                              As shown in Figure~\ref{goals}, solutions found by {an optimization} algorithm are called the 
                                \textbf{Predicted Pareto Front} (PF).           
                               The  list of best solutions of a space found so far is known as the  \textbf{Actual Pareto Front}. As this can be unknowable in practice or prohibitively expensive to generate, it is common to build the actual front  using the union of all optimization outcomes all non-dominated solutions.

\subsection{ Meta-heuristic    Optimizers}
\begin{wrapfigure}{r}{0.56\textwidth}
                \includegraphics[height=4.8cm]{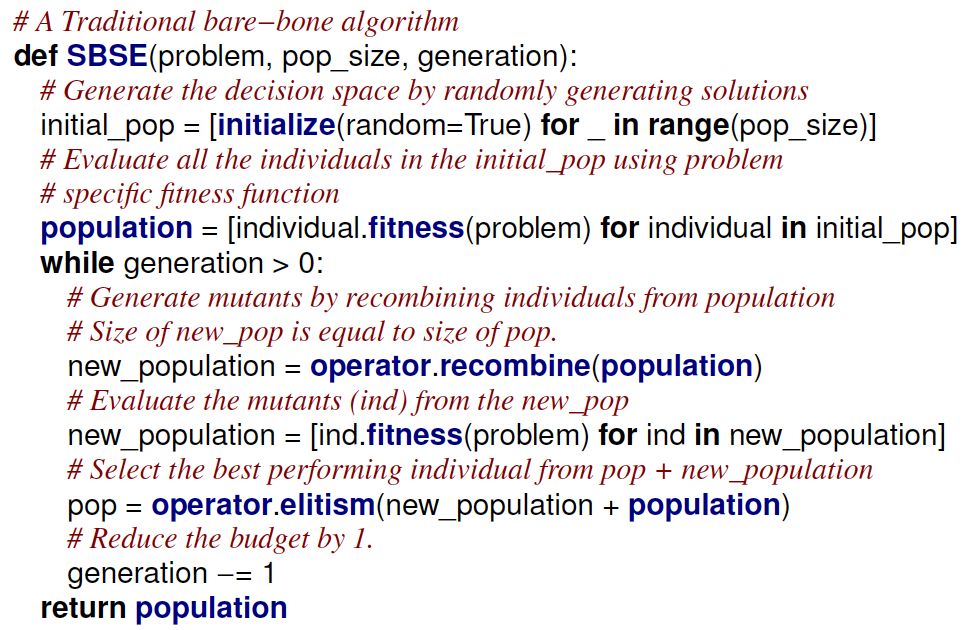}
                \caption{Meta-heuristic search (high-level view).}\label{sbsecode}
                \end{wrapfigure} 
                Any non-trivial software system contains thousands to millions of condition statements (if, case, etc).
Each of these conditions sub-divides the internal state space of a software system.
From a technical perspective, this means that the state space of software
is {\em not} a continuous function (since each conditional creates a different sub-system
with different properties). 
Accordingly, to optimize a software system, simplistic numeric
optimizers are not appropriate. Instead, a common approach is
to use the search-based meta-heuristic methods shown in Figure~\ref{sbsecode}. 
The rest of this section offers notes on that pseudo-code.

               
              Multi-objective optimizers for SE  use   either a model (which represent a software process~\cite{boehm1995cost}) or can be directly applied to any software engineering system (including problems that require evaluating a solution by running a specific benchmark~\cite{krall2015gale}). 
          However the model or software system is created, it can be represented
          in the \textbf{decision space} in a myriad of ways, for example, as Boolean or numerical vectors, as strings, and as graphs.
          (and the space of all solutions that can be represented in 
          this way is the decision space).
          
       Search-based meta-heuristic methods explore and refine a 
        \textbf{population} of candidate
      solutions   using the decision space representation.
                The process of search  typically starts by creating  a
                population of random solutions (valid or invalid)~\cite{saber2017seeding, chen2017beyond, 89chen2018sampling, henard2015combining}.
                 A \textbf{fitness function} then maps the solution (which is represented using numerics) to a numeric scale (also known
                 as \textit{Objective Space}) which is used to distinguish between good and not so good solutions 
                Simply put, the fitness function is a transformation function which converts a point in the decision space to the objective space. 
         
          Some \textbf{variation operator} is then applied to {\bf mutate}
         the solutions (that is, to generate new candidate solutions). Typically, unary operators are called mutation operators, and operators with higher arity are called crossover operators.
         
         Then, it is common to use operators that apply some pressure towards selecting better solutions from the union of the current population and the newly created solutions. This selection is done 
         either deterministically (e.g., elitism operator) or stochastically.         
            An important class of operator is
                \textit{Elitism } which simulates the `survival of the fittest' strategy, i.e., it eliminates not so good solutions thereby preserving the good solutions in the population.

        As shown by the {\tt while} loop of Figure~\ref{sbsecode},
        a meta-heuristic algorithm iteratively improves the population (set of solutions) as, at each iteration, 
          each member of the population might be mutated or eliminated via elitism.
        Each step of this process, which includes generation of new solutions using recombination of the existing population and selecting solutions using the elitism operator, is called a \textbf{generation}. 
        Over successive generations, the population `evolves' (in the best case) toward a globally optimal solution.

\begin{figure}[!t]
\centering%
\includegraphics[height=3.3cm,page=1,trim=160 20 160 20,clip]{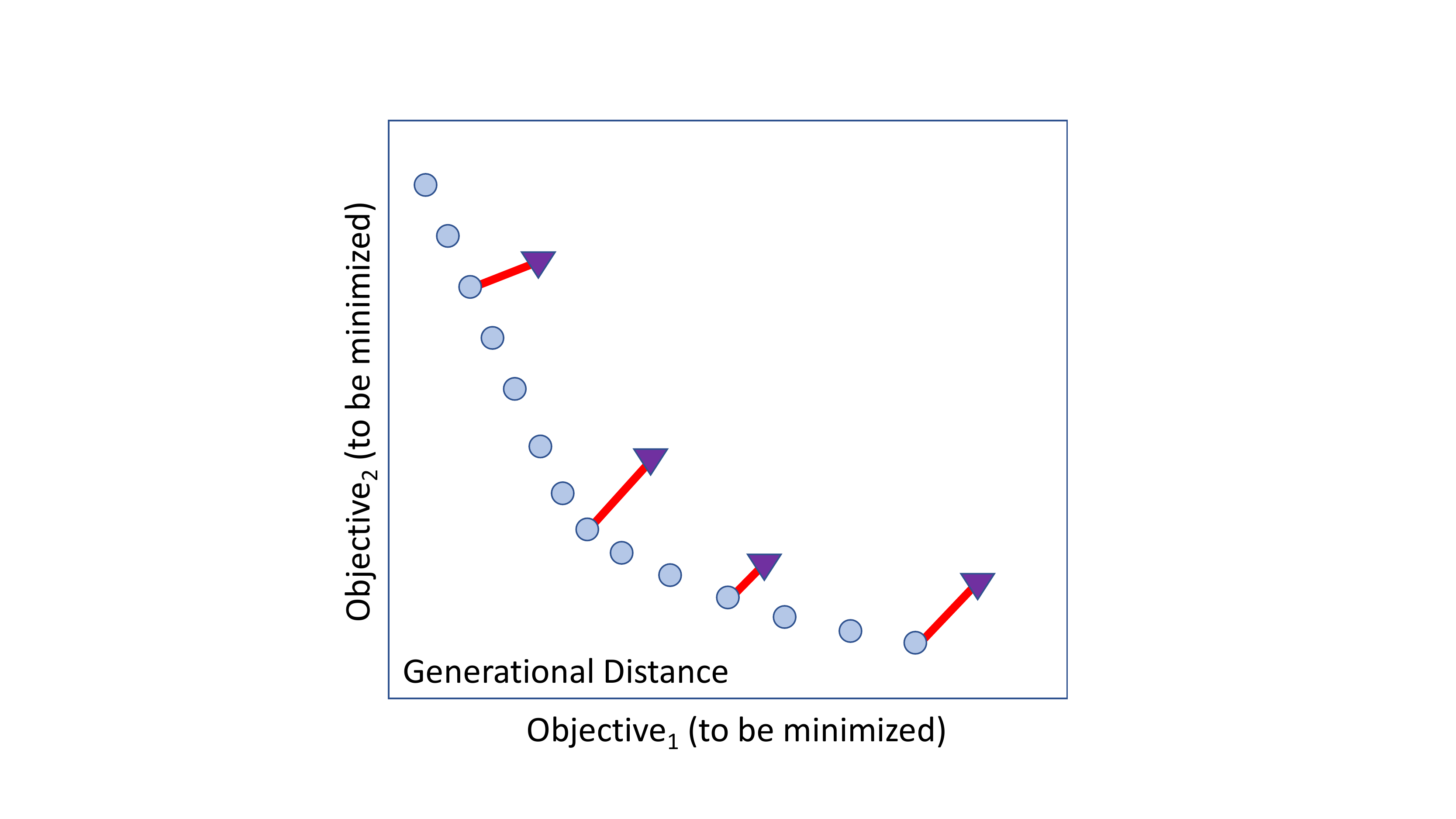}%
\includegraphics[height=3.3cm,page=2,trim=160 20 160 20,clip]{fig/191210duo-indicator-2.pdf}%
\includegraphics[height=3.3cm,page=3,trim=160 20 160 20,clip]{fig/191210duo-indicator-2.pdf} 
\includegraphics[height=3.3cm,page=4,trim=160 20 160 20,clip]{fig/191210duo-indicator-2.pdf}%
\includegraphics[height=3.3cm,page=5,trim=160 20 160 20,clip]{fig/191210duo-indicator-2.pdf}%
   \caption{Optimization evaluation criteria.}\label{fig:definitions}
   \end{figure}                     
   
   \subsection{How is it all assessed?}                             
   For single-objective problems, measures such as \textit{absolute residual} or \textit{rank-difference} can be very useful.
   For   multi-objective problems, the evaluation scores must explore
   trade-offs between (potentially) competing goals.
   \fig{definitions} illustrates some of the standard   evaluation measures used for multi-objective reasoning.
   The rest of this section explains that figure. 
   
   \textbf{Generational distance} is the measure of convergence---how close is the predicted Pareto front to the actual Pareto front. It is defined to measure (using Euclidean distance) how far are the solutions that exist in the population $P$ from the nearest solutions in the Actual Pareto front $A$. In an ideal case, the GD is 0, which means the predicted PF is a subset of the actual PF. Note that it ignores how well the solutions are spread out.
                            
                            \textbf{Spread} is a measure of diversity---how well the solutions in $P$ are spread. An ideal case is when the solutions in $P$ is spread evenly across the Predicted Pareto Front. 
                            
                            \textbf{Inverted Generational Distance} measures both convergence as well as the diversity of the solutions---it measures the shortest distance from each solution in the Actual PF to the closest solution in Predicted PF. Like Generational distance, the distance is measured in Euclidean space. In an ideal case, IGD is 0, which means the Predicted PF is same as the Actual PF.
                            
                            The \textbf{Hypervolume} indicator is used to measure both convergence as well as the diversity of the solutions---hypervolume is the union of the cuboids w.r.t. to a reference point. Note that the hypervolume implicitly defines an arbitrary aim of optimization. Also, it is not efficiently computable when the number of dimensions is large, however, approximations exist.
                            
                         Lastly, we mention \textbf{Approximation:}. The Additive/multiplicative Approximation is an alternative measure which can be computed in linear time (w.r.t. to the number of objectives). It is the multi-objective extension of the concept of approximation encountered in theoretical computer science.

 \BLACK

\section{Claim1: {\RED For software engineering tasks,\BLACK} optimization and data mining are very similar}\label{sec:claim1}
\label{sec:define}
 
 \begin{wrapfigure}{r}{1.5in}
\includegraphics[width=1.5in]{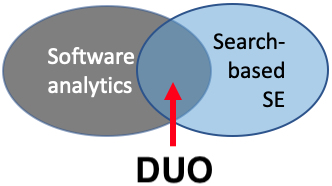}
\caption{About {\IT}.}\label{fig:duodef}
\end{wrapfigure}
At first glance, the obvious connection between data miners and optimizers is that the former build models from data while the latter can be used to exercise those models (in order to find good choices within a model).
 Nevertheless, as far as we can tell, these two areas are currently being explored by different research teams. 
While counter-examples exist,
data miners are used
by {\em software analytics} workers 
and optimizers are used by   researchers into {\em search-based SE} for the most part.
As shown in \fig{duodef},
the field we call ``{\IT}'' combines software engineering work from both  fields.

Search-based software engineering~\cite{harman2012search,5harman2001} characterizes SE tasks as optimizing for (potentially competing) goals; e.g. designing 
a product such that it delivers the {\em most} features at {\em least} cost~\cite{33sayyad2013}.
Software analytics~\cite{bird2015art,27menzies2013,menzim18r}, on the other hand,  is a workflow that distills large amounts of low-value data into small chunks of very high-value data. 
For example, software analytics might build a model predicting where defects might be found in source code~\cite{6gondra2008}. 

For the most part, researchers in these two areas work separately (as witnessed by, say the annual Mining Software Repositories conference and the separate symposium on  Search-based Software Engineering). What will be argued in this paper is that insights and methods from these two fields can be usefully combined.
We say this because  papers that combined data miners and optimizers explore important SE tasks:
\bi
\item
Project management~\cite{13huang2006,19kumar2008,20huang2008,26chiu2007,38minku2013,40sarro2016,42oliveira2010ga,94minku2013analysis,98sarro2012single,89chen2018sampling,31chen2018};
\item Requirements engineering~\cite{100feather2002converging}; 
\item
Software design~\cite{93du2015evolutionary,89chen2018sampling};
\item
Software security~\cite{1ali2018,2sadiq2018};
\item
Software quality~\cite{14zhong2004,35fu2016,44liu2010evolutionary,54agrawal2018better,55de2010symbolic,97sarro2012further,64abdessalem18,47tantithamthavorn2016automated};
\item
Software configuration~\cite{90nair2018finding,87nair2017using,31chen2018};
\item Text mining of software-related
textual artifacts~\cite{43panichella2013effectively,43panichella2013effectively,57agrawal2018wrong,102treude2018per}.
\ei
In our literature review, we have seen four different flavors of this
combined {\IT} approach:

\RED \label{page10}
\bi
\item {\em Mash-ups of data miners and optimizers:} 
In this approach, data miners and optimizers can be seen as separate executables.
For example, 
Abdessalem et al.~\cite{64abdessalem18} generate test cases
for autonomous cars via   a cyclic approach 
where an optimizer reflects on the output of data miners that reflect
on the output of an optimizer (and so on).
\item {\em Data miners acting as optimizers:}
In this approach, there is no separation
between the data miner and optimizer.
For example,
Chen et al.~\cite{89chen2018sampling} show that  their  recursive descent bi-clustering algorithm (which is a data mining technique)
outperforms traditional evolutionary algorithms for the purposes of optimizing SE models.  
\item  {\em Optimizers control the data miners:}
In this approach, the data miner is a subroutine called by the optimizer.
For example, several recent papers improve predictive performance via  optimizers that tune the control parameters of the data miner~\cite{35fu2016,54agrawal2018better,47tantithamthavorn2016automated}.
\item {\em Data miners control the optimizers:} 
In this approach, the optimizer is a subroutine called by the data miner.
For example, Majumder et al.~\cite{88majumder2018500+}
use k-means clustering to divide up a complex text mining problem,
then apply optimizers within each cluster. They report that this method
speeds up their processing by up to three orders of magnitude.
\ei
\BLACK
 To understand why data mining technology is so useful for optimization, and vice versa, we must dive deeper
 into the formal underpinnings of work in this area.
Without loss of generality, an optimization problem is of the following format~\cite{OptDef}:
\begin{equation}
\begin{tabular}{ll}
minimize        & $f_i(\textbf{a})$, $i=1,2,\cdots,n$ \\
subject to      & $g'_j(\textbf{a}) \leq 0$, $j=1,2,\cdots,n'$ \\
                & $g''_k(\textbf{a}) = 0$, $k=1,2,\cdots,n''$ \\
\end{tabular}
\label{eq:optimization-problem}
\end{equation}

\noindent where $\textbf{a} = (a_1,\cdots, a_m) \in \textbf{A}$ is the optimization variable of the problem,\footnote{This definition has been generalized with respect to \cite{OptDef}, not to be restricted to continuous optimization problems, where $\forall a_i, \ a_i \in \rm I\!R$}  $f_i(\textbf{a}): \textbf{A} \rightarrow {\rm I\!R}.$ are the objective functions (goals) to be minimized, $g'_j(\textbf{a})$ are inequality constraints, and $g''_k(\textbf{a})$ are equality constraints.\footnote{The optimization variable is usually identified by the symbol $\textbf{x}$, and the inequality and equality constraints are frequently identified by the symbols $g$ and $h$ in the optimization literature. However, we use the symbols $\textbf{a}$, $g$ and $g''$ here to avoid confusion with the terminology used in data mining, which is introduced later in this section.} Sometimes, there are no constraints (so $n'=0$ and $n''=0$). Also, we say a {\em multi-objective problem} has $n>1$ objectives, as opposed to a {\em single-objective problem}, where $n=1$.

One obvious question about this general definition is ``where do the functions $f,g',g''$ come from?''. Traditionally, these have been built by hand but,
as we shall see below, $f,g',g''$ can be learned via data mining. That is, optimizers can {\em explore}
the functions {\em proposed} by a learner.

An example of an optimization problem in the area of software engineering is to find a subset $\textbf{a}$ of requirements that maximizes value $f(\textbf{a})$ if implemented\footnote{Any maximization problem can be re-written as a minimization problem.}, given a constrained budget $g'_0(\textbf{a}) \leq 0$, where $g'_0(\textbf{a}) = cost(\textbf{a}) - budget$ \cite{Sagrado2011}.
Many different algorithms exist to search for solutions to optimization problems. Table \ref{tbl:optimizers} shows the optimization algorithms that have been used by the software engineering community when applying DUO.

Data mining is a problem that involves finding an approximation $\hat{h}(\textbf{x})$ of a function of the following format:
\begin{equation}\textbf{y}=h(\textbf{x})\label{eq:learning}\end{equation} 

\noindent where $\textbf{x} =(x_1,\cdots, x_p) \in \textbf{X}$ are the input variables, $\textbf{y} = (y_1,\cdots, y_q) \in \textbf{Y}$ are the output variables of the function $h(\textbf{x}): \textbf{X} \rightarrow \textbf{Y}$, \textbf{X} is the input space and \textbf{Y} is the output space. The input variables $\textbf{x}$ are frequently referred to as the independent variables or input features, whereas $\textbf{y}$ are referred to as the dependent variables or output features.  

An example of a data mining problem in software engineering is software defect prediction~\cite{hall2012systematic}. Here, the input features could be a software component's size and complexity, and the output feature could be a label identifying the component as defective or non-defective. Many different machine learning algorithms can be used for data mining.
\tbl{learn} shows data mining  algorithms  used by the software engineering community when applying DUO.

The functions $h(\textbf{x})$ and $\hat{h}(\textbf{x})$ do \textit{not} necessarily correspond to the optimization functions $f_i(\textbf{a})$ depicted in Eq.~\ref{eq:optimization-problem}. However, the true function $h(\textbf{x})$ is unknown. Therefore, data mining frequently relies on machine learning algorithms to learn an approximation $\hat{h}(\textbf{x})$ based on a set $D = \{(x_i,y_i)\}_{i=1}^{|D|}$ of known examples (data points) from $h(\textbf{x})$. And, learning this approximation typically consists of searching for a function $\hat{h}(\textbf{x})$ that minimizes the error (or other predictive performance metrics) on examples from $D$. Therefore, learning such approximation \textit{is} an optimization problem of the following format:
\begin{equation}
\begin{tabular}{ll}
minimize        & $f_i(\textbf{a})$, $i=1,\cdots,n$ \\
\end{tabular}
\label{eq:learning-problem}
\end{equation}

\noindent where $\textbf{a} = (\hat{h}(\textbf{x}), D)$, and $f_i$ are the predictive performance metrics obtained by $\hat{h}(\textbf{x})$ on $D$. The functions $f_i(\textbf{a})$ depicted in Eq.~\ref{eq:learning-problem} thus correspond to the functions $f_i(\textbf{a})$ depicted in Eq.~\ref{eq:optimization-problem}. An example of performance metric function would be the mean squared error, defined as follows:
\begin{equation}
f(\hat{h}(\textbf{x}), D) = \frac{1}{|D|} \sum_{(x_i,y_i) \in D} (y_i -  \hat{h}(x_i))^2
\label{eq:mae}
\end{equation}

As we can see from the above, solving a data mining problem means solving an optimization problem, i.e., optimization and data mining are very similar. Indeed, several popular machine learning algorithms \textit{are} optimization algorithms. For example, gradient descent for training artificial neural networks, quadratic programming for training support vectors machines and least squares for training linear regression are optimization algorithms.

From the above, we can already see that optimization is of interest to data mining researchers, even though this connection between the two fields is not always made explicit in software engineering research. More explicit examples of how optimization is relevant to data mining in software engineering include the use of optimization algorithms to tune the parameters of the data mining algorithms, as mentioned at the beginning of this section and further explained in Sections  \ref{sec:claim3} and \ref{sec:claim5}. We consider such more explicitly posed connections between data mining and optimization as a form of DUO.


A typical distinction made between the optimization and data mining fields is data mining's need for generalization. Despite the fact that data mining uses machine learning to search for approximations $\hat{h}(\textbf{x})$ that minimize the error on a given dataset $D$, the true intention behind data mining is to search for approximations $\hat{h}(\textbf{x})$ that minimize the error on unseen data $D'$ from $h(\textbf{x})$, i.e., being able to generalize. As $D'$ is unavailable for learning, data mining has to rely on a given known data set $D$ to find a good approximation $\hat{h}(\textbf{x})$. Several strategies can be adopted by machine learning to avoid poor generalization despite the unavailability of $D'$. For instance, the performance metric functions $f_i(\hat{h}(\textbf{x}), D)$ may use regularization terms \cite{Bishop}, which encourage the parameters that compose $\hat{h}(\textbf{x})$ to adopt small values, making $\hat{h}(\textbf{x})$ less complex and thus generalize better. Another strategy is early stopping \cite{Bishop}, where the learning process stops early, before finding an optimal solution that minimizes the error on the whole set $D$. 


However, even the distinction above starts to become blurry when considering that generalization can also frequently be of interest to optimization researchers. 
For example, in software configuration optimization (Section  \ref{sec:claim4}), the true optimization functions $f_i(\textbf{a})$ are often too expensive to compute, requiring machine learning algorithms to learn approximations of such functions. Optimization functions approximated by machine learning algorithms are referred to as surrogate models, and correspond to the approximation $\hat{h}(\textbf{x})$ of Eq.~\ref{eq:learning-problem}. These approximation functions are then the one optimized, rather than the true underlying optimization function. Even though an optimization algorithm to solve this problem does not attempt to generalize, a data mining technique can do so on its behalf. We consider this as another form of DUO. 
Sections \ref{sec:claim3} and \ref{sec:claim4} explain how several other examples of software engineering problems are indeed both optimization and data mining problems at the same time, and how different forms of DUO can help solving these problems.

Overall, this section shows that optimization and data mining are very similar to each other, and that the typical distinction made between them can become very blurry when considering real world problems. Several software engineering problems require both optimization and generalization at the same time. Therefore, many of the ideas independently developed by the field of optimization are applicable to improve the field of data mining, and vice-versa. Sections \ref{sec:claim3} to \ref{sec:claim5} explain how useful DUO can and could be.

\section{Claim2: {\RED For software engineering tasks,\BLACK}  optimizers can greatly improve data miners}\label{sec:claim3}

\begin{figure}
\includegraphics[width=4.8in]{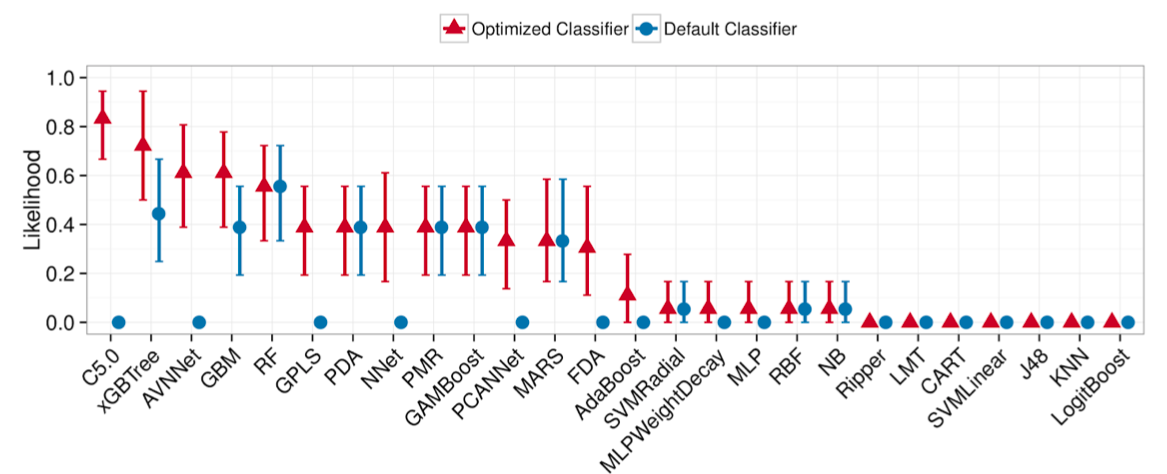}
\caption{Effects of hyperparameter optimization on control parameters of  a learner from~\cite{47tantithamthavorn2016automated}. Blue dots and red triangle show the mean performance before and after tuning (respectively). 
X-axis shows different learners. Y-axis shows the frequency at which a learner was selected to be ``top-ranked'' (by a statistical analysis). Vertical lines show the variance of that selection process over repeated runs.}\label{fig:tanit}
\end{figure}

One of the most frequent ways to integrate
data mining and optimization is via {\em hyperparameter optimization}. This is the art of tuning the parameters that control the choices within a data miner.
While these can be set  manually\footnote{Using a process called ``engineering judgement''; i.e. guessing.} we found that several papers
in our   literature review used optimizers to
 automatically find the best parameters~\cite{54agrawal2018better,%
35fu2016,%
44liu2010evolutionary,%
97sarro2012further,%
14zhong2004,%
102treude2018per,
42oliveira2010ga}. 
There are many reasons why this is so:

\bi
\item  These control parameters are many and varied. 
Even supposedly simple algorithms come with a daunting number of options. For example, the scitkit-learn toolkit lists over a dozen configuration options for Logistic Regression\footnote{
\href{https://scikit-learn.org/stable/modules/generated/sklearn.linear_model.LogisticRegressionCV.html}{https://scikit-learn.org/stable/modules/generated/sklearn.linear\_model.LogisticRegressionCV.html}, accessed 30 November 2018.}. This is an important point since recent research shows that the
{\em more} settings we add to software, the {\em harder} it becomes for humans to use that software~\cite{Xu:2015}.
\item Manually fine-tuning these parameters to obtain the best results for a new data set, is not only tedious, but also can be biased by a human's (mis-)understanding of the inner workings of the data miner.
\item The hurdle to implement or apply a ``successful'' heuristic for automated algorithm tuning is low since (a) the default settings are often not optimal for the situation at hand, and (b) a large number of optimization packages are readily available~\cite{rainville2012deap,Durillo2011}.
Some data mining tools now  come 
with built-in optimizers or tuners; e.g the SMAC implementation built into
the latest versions of Weka~\cite{hutter2011sequential,Hall:2009};
or the CARET package in ``R''~\cite{JSSv028i05}.
\item
Several results report spectacular improvements 
in the performance of data miners after tuning~\cite{54agrawal2018better,%
35fu2016,%
44liu2010evolutionary,%
97sarro2012further,%
47tantithamthavorn2016automated,%
14zhong2004,%
102treude2018per}.
\ei
As evidence to the last point, we offer two examples.
Tantithamthavorn et al.~\cite{47tantithamthavorn2016automated}
applied the CARET  grid search~\cite{JSSv028i05} to improve the predictive performance of classifiers. Grid search is an exhaustive search 
across a pre-defined set of hyperparameter values. It is implemented as a set of for-loops, one for each hyperparameter. Inside the inner-most for-loop,
some learner is applied to some data to assess
the merits of a particular set of hyperparameters.  
Based on statistical methods, Tantithamthavorn et al. ranked all the learners in their study to find the top-ranked tuned learner. As shown in  \fig{tanit},
there is some variability in the likelihood of being top-ranked
(since their analysis was repeated for multiple runs).

\begin{figure}
\includegraphics[width=2.4in]{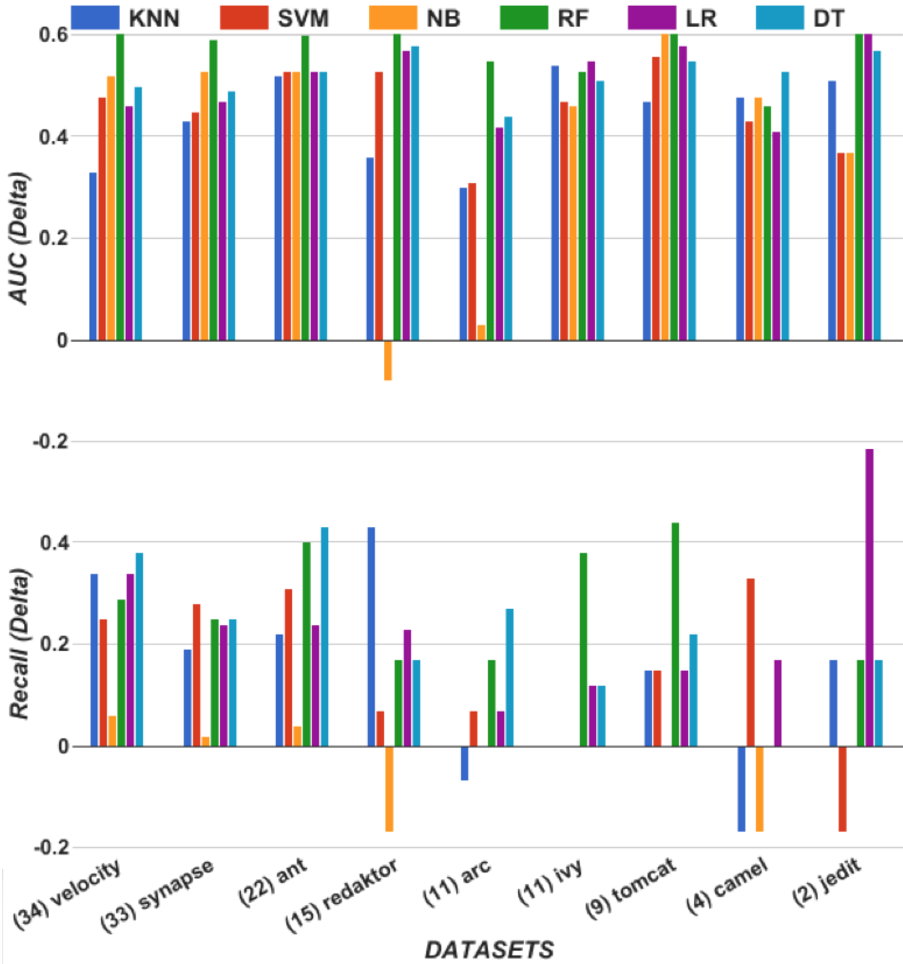}\includegraphics[width=2.4in]{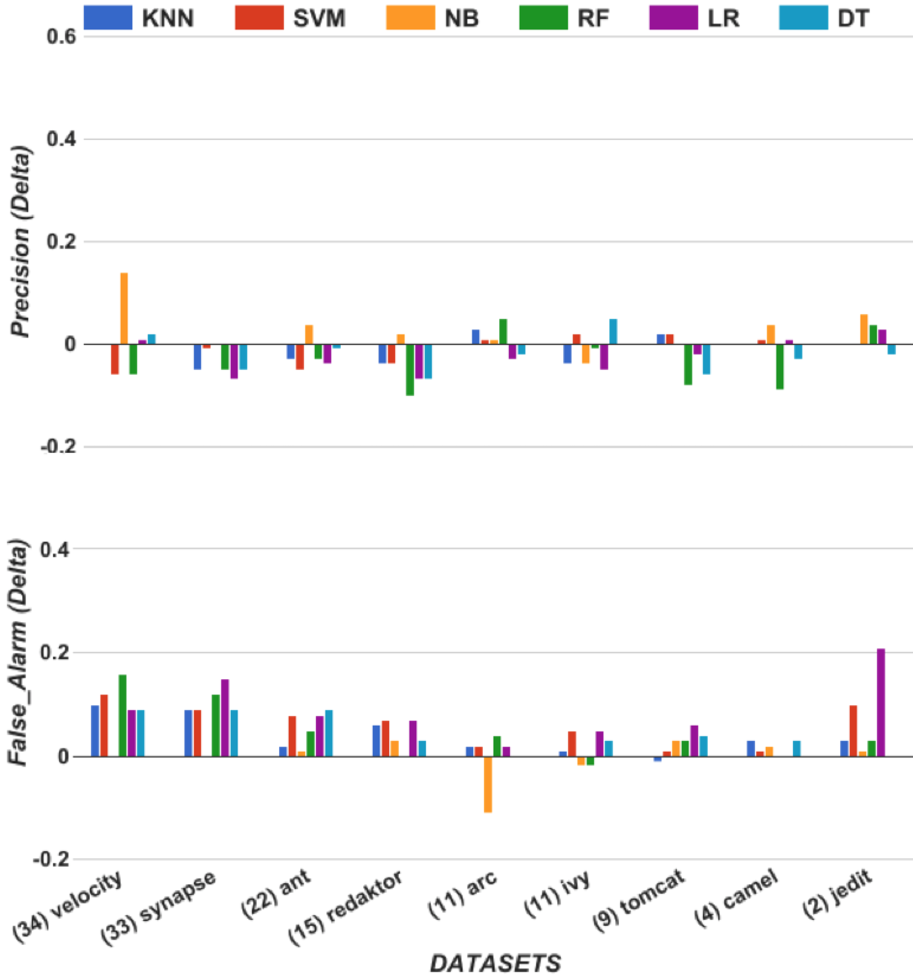}
\caption{Effects of hyperparameter optimization on control parameters of a data pre-processor from~\cite{54agrawal2018better}. Different colored vertical bars refer to different learners:
KNN=nearest neighbor; SVM= support vector machine; NB=naive Bayes; RF=random forest; LR=logistic regression; DT=decision trees. 
X-axis shows different datasets. Y-axis   shows
the {\em after - before} values of four different performance scores: recall, precision, false alarm and AUC (area under the false positive vs true positive rate). For false alarms,
{\em better} deltas are {\em smaller}. For all other measures, the {\em larger} the {\em better}.}\label{fig:aucRecall}
\end{figure}
\noindent From \fig{tanit} we can see that:
\bi
\item Hyperparameter optimization {\em never makes performance worse}.
 We say this since the red triangles (tuned results) are never lower than their blue dot counterpart (untuned results). 
 \item
 Hyperparameter  optimization is {\em associated with some spectacular performance improvements}. Consider the first six left-hand-side blue dots at $Y=0$. These
 show  all untuned learners that are never   top-ranked. After tuning, however, the ranking of these learners is very different. Note once we tune two of these seemingly ``bad''
learners (C5.0 and AVNNet), they become part of the top three best learners.
\ei
For another example of the benefits of hyperparameter optimization, consider 
\fig{aucRecall}. This work shown by Agrawal et al.~\cite{54agrawal2018better} where differential evolution  tuned a data pre-processor called SMOTE~\cite{chawla2002smote}. SMOTE rebalances training data by discarding items of the majority class and synthesizing artificial members of the minority class. As discussed in Table~\ref{tbl:optimizers}, differential evolution~\cite{Storn:1997} is an evolutionary algorithm that uses a unique mutator that builds new candidates by combining
three older candidates.
From \fig{aucRecall}  we can observe the same patterns seen in \fig{tanit}:
\bi
\item
 Hyperparameter optimization {\em rarely makes performance much worse}. There are some losses in precision and false alarms grow slightly. But overall, these changes are very small.
 \item
 Hyperparameter optimization is {\em associated with some spectacular performance improvements}.
 The improvements in recall can be large and the improvements in AUC are the largest
 we have ever seen of any method, ever, in the software analytics literature.
\ei

Another advantage of hyperparameter optimizers is that it can tune learners such that they succeed on multiple criteria. Standard learners have their goals ``hard-wired'' (e.g. minimize entropy).  This can be a useful design choice since it allows algorithm designers to produce very fast systems that scale to very large data sets.  That said, there are many   situations where the business users have {\em multiple} competing goals; e.g. deliver {\em more} functionality,
in {\em less} time, with {\em fewer} bugs. 
For further examples of multiple business goals in software analytics, see Table 23.2 of~\cite{menzies2013data}.

While the goals of data miners are often hard-wired, optimizers
can accept multiple goals as part of the specification of a problem (see the $g$ terms within Equation~\ref{eq:optimization-problem}).
This means optimizers can explore a broader range of goals than data miners. For example: 
\bi
\item
 Minku and Yao \cite{MinkuYao2013} used multi-objective evolutionary algorithms to generate neural networks for software effort estimation, with the objectives of minimizing different error metrics.
 \item
 Sarro et al.~\cite{40sarro2016} used multi-objective genetic programming for software effort estimation, with the objectives of minimizing the error and maximizing the confidence on the estimates.
\ei
 An interesting variant of the above is optimization (in the form of reinforcement learning)  is model selection over time \cite{MinkuYao2014,MinkuYao2017}. Depending on the problem being tackled, the best predictive model to be used for a given problem may change over time, due to changes suffered by the underlying data generating process of this problem. For such problems, model choice has to be continuously performed over time, rather than performed only once, prior to model usage. To deal with that in the context of software effort estimation, Minku and Yao \cite{MinkuYao2014,MinkuYao2017} monitor the predictive performance of software effort estimation models created using software effort data from different companies over time. This predictive performance was monitored based on a time-decayed performance metric derived from the reinforcement learning literature, computed over their effort estimations provided for a given company of interest over time. The models whose predictive performances are recently the best are then selected to be emphasized when performing software effort estimations to the company of interest. When combined with transfer learning \cite{MinkuYao2014}, this strategy enabled a reduction of 90\% in the number of within-company effort data required to perform software effort estimation, while maintaining or sometimes even slightly improving predictive performance. This is a significant achievement, given that the cost of collecting within-company effort data is typically very high.

\subsection{A Dozen  Tips for Using Optimizers}\label{sec:12tipso}

The next section describes some of the
problems associated with using optimizers. Before
that, this section offers some rules of thumb for
software engineers wishing to use 
{\em optimizers}
in the manner recommended by this paper.

Just to say the obvious,
we cannot {\em prove} the utility of
the following heuristics. That said,
when we work with our industrial
partners or graduate students, we   often say the following.

1. Start with  a clear and detailed problem formulation. If you do not understand the problem well, then  the proposed approach to solve the problem may not deliver what is desired.

2. Visualize first, optimize second.
That is, try to visualize the trade-offs between objectives before going into optimization
by (e.g.) randomly generating some candidate solutions and plotting the generated
performance scores (and for multi-goal reasoning, visualize the principal components
of the objective space). We say this since (sometimes) glancing at such a visualization
can lead to insights such as ``this problem divides into two separate problems that we should
explore separately''.

3. Optimizers of the kind explored
here are found in many open source toolkits
written in various languages (e.g. JAVA,
C++, Python) such as 
  jMETAL~\cite{Durillo2011}
or PAGMO/PyGMO (esa.github.io/pagmo2).

4. No optimizer works best for all applications~\cite{Wolpert97}.
That said, once it can be shown 
that one optimizer is better than another
then the set of better optimizers is exponentially smaller~\cite{Montanez13}.
Hence, when faced with a new problem, it is useful to 
try several optimizers and stop after two significant improvements have been achieved over some initial baseline result.
In terms of algorithms, 
a useful set to try first are NSGA-2~\cite{deb02} (since everyone tries this one), MOEA/D~\cite{zhang2007moea} (since it is fast), and differential evolution~\cite{Storn:1997}
 (since it is so simple to use).

5. Further to the last point, any industrial application of this paper should
try several optimizers. For that purpose, it is useful to apply faster optimizers
(e.g. MOEA/D)
before slower methods (e.g. NSGA-II).  Note that if your preferred method is very
slow, then it can be speeded up via data mining (see point\#1).

6. As said in \S\ref{sec:claim3},
if optimizers run too slowly, 
use data miners to divide the data then
and apply optimization
to each segment~\cite{88majumder2018500+}.

7. 
If your users cannot understand what the optimizer is saying, use data mining to produce a summary of the results.

8.  Watch out for changes in the problem over time -- they may cause a previous optimal solution to become poor.

9. More specifically, insights can change with the computational budget. That is, conclusions that seem most useful after 1,000 evaluations might be superseded by the results from 5,000 evaluations on.
Hence, if possible, before reporting a conclusion to users, try doubling the number of evaluations to see if your current conclusions still hold.

10. True multi-objective formulations are less biased than linear combinations of objectives.
We say this since, sometimes, it is suggested
to reduce a multi-optimization problem to a simpler
single-objective problem by adding ``magic weights'' to each objective (e.g. ``three times
the speed of the car plus twice times the cost of the car''). Such ``magic weights''
introduce an unnecessary bias to the analysis. These magic weights can be avoided by using
a true multi-objective algorithm (e.g. NSGA-II, MOEA/D, differential evolution (augmented with 
Algorithm~\ref{cdom}).

11. When optimizing for one or two goals,
a simple predicate is enough to select which
solution is better (specifically: $x$ is not worse 
than $y$ on both objectives; and $x$ is better on at least one).

12.
But when optimizing for more goals,
Zitzler's indicator method~\cite{Zitzler2004IndicatorBasedSI} 
might be needed~\cite{33sayyad2013}.
This indicator method was shown in Algorithm~\ref{cdom}.

13.
Finally, for multiple goals, sometimes it is useful to focus first
on a small number of most difficult goals (then use the results
of that first study to ``seed'' a second study that explores
the remaining goals)~\cite{Sayyadzz}.




\subsection{Problems with Hyperparameter Optimization}
In summary, optimization is associated with some spectacular improvements in data mining. Also,   by applying
optimizers to data miners, they can better address the  domain-specific and  goal-specific queries
of different users.

One pragmatic drawback with hyperparameter optimization is its    associated
runtime. Each time a new hyperparameter setting is evaluated, a learner must be called on some training data, then tested on some separate ``hold-out'' data. This must be repeated, many times. In practice, this can take
a long time to terminate:
\bi
\item
When replicating 
the  Tantithamthavorn et al.~\cite{47tantithamthavorn2016automated}  experiment, Fu et al.~\cite{fu2016differential}
implemented tuning using grid search and differential evolution. 
That study used   20 repeats for tuning random forests (as the target learner), and optimized four different measures
of AUC, recall, precision, false alarm, that grid search required 109 days of CPU.
Differential evolution and grid search required $10^4$ and $10^7$ seconds to terminate,
respectively\footnote{Total time to process  20 repeated runs across multiple subsets of the data, for multiple data sets.}.
\item   Xia et al.~\cite{Xia2018HyperparameterOF} reports experiments with hyperparameter optimization for software effort estimation. In their domain, depending on what data set was processed, it took
 140 to 700 minutes  (median to maximum) to compare seven ways to optimize two data miners. If that experiment is repeated 30 times for statistical validity, then the full experiment would take 70 to 340 hours (median to max).

\ei
While the above runtimes might seem practical to some researchers,
we note that other hyperparameter optimization tasks take a very long time to terminate.  Here are the two worst (i.e. slowest) examples
that we know of, seen in the recent SE literature:
\bi
\item E.g.    decades of CPU time were needed by Treude et al.~\cite{102treude2018per} to achieve a 12\% improvement over the default settings;
\item E.g.   15 years of CPU were needed in the hyperparameter optimization of software clone detectors by Wang et al.~\cite{wang2013searching}.
\ei
One way to address these slow runtimes is via (say) cloud-based CPU farms.  Cross-validation experiments can be easily parallelized just by running  each cross-val on a separate core. But 
the cumulative costs of that approach can be large.  For example, recently while developing a half million US dollar research proposal, we 
estimate how much it would cost to run the same kind of hardware as seen in related work. For that three year project,  two graduate students could easily use  \$100,000 in CPU time -- which is a large percentage of the grant (especially since, after university extracts its overhead change, that \$500K grant would effectively be \$250K).

Since using optimizers for hyperparameter optimization can be very resource-intensive, the  next section discusses {\IT} to significantly reduce that cost.

\section{Claim3: {\RED For software engineering tasks, \BLACK}  data miners can greatly improve optimization}\label{sec:claim4}

  
The previous section mentions the benefits of optimizers for data mining, but warned that such optimization can be  slow. One way to speed up  optimization  is to divide the total problem into many small sub-problems. As discussed in this section, this can be done using data mining. That is, data mining can be used to optimize optimizers. 

Another benefit of data mining is that, as discussed below, it can generalize and summarize the results of optimization. That is, data mining can make optimization results more comprehensible.

\subsection{Faster Optimization}\label{tion:fastop}

It can be a very simple matter to implement data miners improving optimizers. Consider, for example,     Majumder et al.~\cite{88majumder2018500+} who were looking for the connections between posts to StackOverflow (which is a popular online question and answer forum for programmers):

\bi
\item
An existing deep learning approach~\cite{xu16} to that problem was so slow that it was hard to reproduce prior results. 
\item
Majumder et al. found that they could get equivalent results 500 to 900 times faster (using one core or eight cores working in parallel) just by applying k-means clustering to the data, then running their  hyperparameter optimizer (differential evolution) within each cluster.
\ei
Another example of data mining significantly improving optimization, consider the  {\em sampling} methods of Chen et al.~\cite{31chen2018,89chen2018sampling}. This team explored optimizers for a variety of (a)~software process models as well as the task of (b)~extracting products from a product line description. These are multi-objective problems that struggle to find solutions that (e.g.) minimize development cost while maximizing the delivered functionality (and several other goals as well). The product extracting task was particularly difficult. Product lines were expressed as trees of options with ``cross-tree constraints'' between sub-trees. These constraints mean that decisions in one sub-tree have to be carefully considered, given their potential effects on decisions in other sub-trees. Formally, this makes the problem NP-hard and in practice, this product extraction process was known to defeat state-of-the-art theorem provers~\cite{pohl11}, particularly for large product line models (e.g. the ``LVAT'' product line model of a LINUX kernel contained 6888 variables within a network of 343,944 constraints~\cite{89chen2018sampling}).

Chen et al. tackled this optimization problem using data mining to look at just a small subset of the most informative examples.
Chen et al. call this approach a ``sampling'' method.
\BLACK
Specifically, they used
a recursive bi-clustering algorithm over a large initial population to isolate the superior candidates.
As shown in the following list, this  approach is somewhat different to the more standard
genetic algorithms approach:

\bi
\item Genetic algorithms (in SE) often start with a population of $N=10^2$ individuals. 
\item On the other hand, samplers start with a much larger population of $N=10^4$ individuals.
\item Genetic algorithms run for multiple {\em generations} where useful variations of individuals in generation $i$ are used to seed generation $i+1$. 
\item On the other hand, samplers run for a single generation, then terminate.
\item Genetic algorithms evaluate all $N$ individuals in all generations.
\item On the other hand, the samplers of Chen et al. evaluate pairs of distant points. If one point proves to be inferior then it is pruned along with all individuals in that half of the data. Samplers then recursively prune the surviving half. In this way, samplers only evaluate $O(2{\log}_2N)$ of the population,
\ei
Regardless of the above differences, the goal of genetic algorithms and samplers is the same:  find options that best optimize some competing set of goals. In comparisons with NSGA-II (a widely used genetic algorithm~\cite{deb02}), Chen et al.'s sampler usually optimized the same, or better, as the genetic approach. Further, since samplers only evaluate $O(2{\log}_2N)$ individuals, sampling's median cost was just 3\% of runtimes and 1\% of the number of model evaluations (compared to only running the genetic optimizer)~\cite{89chen2018sampling}.

For another example of data miners speeding up optimizers, see the work of Nair et al.~\cite{87nair2017using}. That work characterized the software configuration optimization problem as ranking a (very large) space of configuration options, without having to run tests on all those options. For example, such configuration optimizers might find a parameter setting to SQLite's configuration files that maximized query throughput. Testing each configuration requires re-compiling the whole system, then re-running the entire test suite. Hence, testing the three million valid configurations for SQLite is an impractically long process.

The key to quickly exploring such a large space of options, is  {\em surrogate modeling}; i.e. learning an approximation to the response variable being studied. The two most important properties of such surrogates are that they are much faster to evaluate than the actual model, and that the evaluations are precise. Once this approximation is available then configurations can be ranked by generating estimates from the surrogates. Nair et al. built their surrogates using a data miner; specifically, a regression tree learner called CART~\cite{breiman84}. An initial tree is built using a few randomly selected configurations. Then, while the error in the tree's predictions decreases, a few more examples are selected (at random) and evaluated. Nair et al. report that this scheme can build an adequate surrogate for SQLlite after 30-40 evaluated examples. 

For this paper, the key point of the Nair et al. work is that this data mining approach scales to much larger problems then what can be handled via standard optimization technology. For example, the prior state-of-the-art result in this area was work by Zuluga et al.~\cite{Zuluaga:2013} who used a Gaussian process model (GPM) to build their surrogate. GPMs have the advantage that they can be queried to find the regions of maximum variance (which can be an insightful region within which to make the next query). However, GPMs have the disadvantage that they do not scale to large models. Nair et al. found that the data mining approach scaled to models orders of magnitude larger than the more standard optimization approach of Zuluga et al.

\subsection{Better Comprehension of User Goals}\label{sect:betterco}


Aside from speeding up optimization, there are other benefits of adding data miners to optimizers. 
If we combine data miners and optimizers then we can (a)~better understand user goals to (b)~produce results that are more relevant to our clients.

To understand this point, we first note that modern data miners run so quickly since they are highly optimized to achieve a single goal (e.g. minimize class  entropy or variance). But there are many situations where the business users have {\em multiple} competing goals; e.g. deliver {\em more} functionality,
in {\em less} time, with {\em fewer} bugs. 
A standard data miner (e.g. CART) can be kludged to handle multiple goals reasoning, as follows: compute the class attribute via some {\em aggregation function} that uses some ``magic weights'' $\beta$, e.g., $\mathit{class\;\textunderscore value}= \beta_1\times\mathit{goal1} + \beta_2\times\mathit{goal2} + ...$
But using an aggregate function for the class variable is a kludge, for three reasons.
Firstly, when users change their preferences about $\beta_i$, then the whole inference must be repeated.

Secondly,
the $\beta_i$ goals may be inconsistent and conflicting. A repeated result in decision theory is that user preferences may be nontransitive~\cite{Fishburn1991} (e.g. users rank $\beta_1<\beta_2$ and $\beta_2<\beta_3$ but also $\beta_3<\beta_1$). Such intransitivity means that a debate about how to set $\beta$ to a range of goals may never terminate.

\begin{figure} 
\centering
\rotatebox{90}{\hspace{24mm} \tiny\textsf{value}}
\includegraphics[width=2.5in,trim=20 20 0 0,clip,]{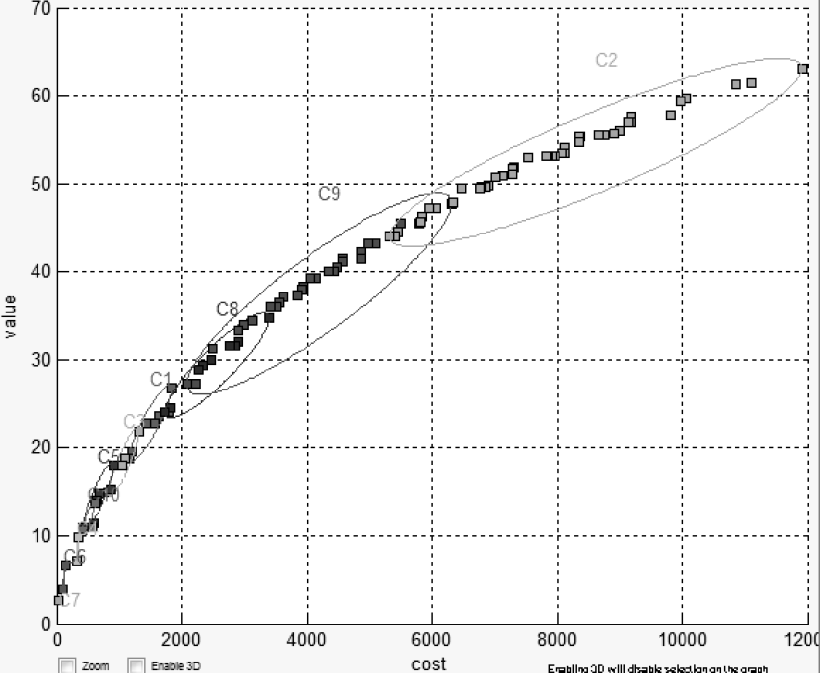}\\
\vspace{-1.5mm}\hspace{4mm}{\tiny\textsf{cost}}
\caption{Clusters of the solution frontier. Frontier generated by reasoning over the goals (in this case, {\em minimize} cost and \emph{maximize} value). Clusters generated by reasoning over the decisions that lead to those goals. From~\cite{Veerappa11}.}\label{fig:veer}
\end{figure}

Thirdly, rather than to restrict an inference to the whims of one user, it can be insightful to let an algorithm generate solutions across the space of possible preferences. This approach was used by Veerappa et al.~\cite{Veerappa11} when exploring the requirements of the London Ambulance system. They found that when they optimized those requirements, the result was a {\em frontier} of hundreds of solutions like that shown in \fig{veer}. Each member of that frontier is   trying to push out on all objectives (and  perhaps failing on some). Mathematicians and economists call this frontier the {\em Pareto frontier~\cite{pareto1906manuale}}. Others call it the {\em trade-off space} since it allows users to survey the range of compromises (trade-offs) that must be made when struggling to achieve multiple, possibly competing, goals.

\fig{veer} is an illustrative example of how data mining can help optimizers. In that figure, the  results are grouped by a data miner (a clusterer). The decisions used to reach the centroid of each cluster are a specific example, each with demonstrably different effects. When showing these results to users, Veerappa et al. can say (e.g.) ``given all that you've told us, there a less than a dozen kinds of solutions to your problem''. That is:
\bi
\item
Step\#1: Optimize: here are the possible decisions;
\item
Step\#2: Data mine: here is a summary of those decisions;
\item Step\#3: Business users need only debate the options in the summary.
\ei


Note that \fig{veer} lets us comment on the
merits of global vs local reasoning. {\em Global} reasoning might return the average
properties of all the black points in that figure. But if we apply {\em local} reasoning, we can find specialized
models within each cluster of  \fig{veer}. The merits
of local vs global reasoning are domain-specific but in the
specific case of \fig{veer}, there seem to be some key differences between the upper right and lower left clusters.

Another approach to understanding trade-off space is contrast-rule generation
as done by (e.g.) Menzies et al.'s STAR algorithm~\cite{Menzies:2007}.
STAR divided  results into a 10\% ``best'' set and a 90\% ``rest'' set. A Bayesian contrast set procedure  ranks all $N$ decisions in descending order (best to worst). STAR then re-runs the optimizer $i\in N$ times, each time pre-asserting the first $i$-the ranges.
 For example,
suppose 10,000 evaluations lead to  $B=1000$ `best' solutions and $R=9000$ rest.
\bi
\item
Let $a$ be ``analyst capability''  and  let ``$a$=high''  appear 50 times in best and 90 times in rest.
\item
Let 
$u$ be ``use of software tools'' and let ``$u$=high`` appear 100 times in best and 180 times in rest.
So 
$b_{a.high}= 50/B = 50/1000= 0.05$;
$r_{a.high}= 90/R = 90/9000=0.01$;  
 $b_{u.high}=100/1000=0.1$; and
$r_{u.high}=180/9000=0.02$.
 \item
 STAR sorts ranges via $s=b^n/(b+r)$, where $n$ is a constant
 used to reward ranges with high support in ``best''. 
E.g. if  $n=2$ then
  $s_{a.high}=0.042$ and $s_{u.high}=0.083$.
 That is,  STAR thinks that the high use of software tools  is more important than 
  high analyst capability.
  \ei
  \BLACK
  The output of STAR result is a graph showing the effects of taking the first best decision, the first two best decisions, and so on. In this way, STAR would report to users a succinct rule set advising them what they can do if they are willing to change just one thing, just two things, etc~\cite{Menzies:2007}.

Before ending this section, we stress the following point: 
{\em it can be very simple to add a data miner to an optimizer}.
For example:
\bi
\item
STAR's contrast set procedure  described above, is very simple to code (around 30 lines of code in Python).  
\item
Recall from the above that Majumder et al. speed up their optimzer by 500 to 900 times, just by prepending a k-means clusterer to an existing optimization process. 
\ei

\subsection{A Dozen   Tips for Using Data Mining}\label{sec:tipsdm}

This section offers some rules of thumb for
software engineers wishing to use 
{\em data miners}
in the manner recommend by this paper.

As said above, just to say the obvious,
we cannot {\em prove} the utility of
the following heuristics. That said,
when we work with our industrial
partners or graduate students, we   often say the following.

1. Data miners of the kind explored here are found in many open source toolkits written
in various languages (e.g. JAVA, Python) such a WEKA~\cite{Hall:2009}  or Scikit-Learn~\cite{scikit-learn}.

2. If your data  mining problem has many goals,
consider replacing your data mining algorithms
with an optimizer.

3. Avoid the use of the off-the-shelf parameters. Instead,
use optimizers to select better settings for the local
problem. For more on this point, see {\bf Claim4}, below.

4. Check for conclusion stability. Once you make a conclusion, repeat the entire process ten times using a 90\% random sample of the data each time.
Do not tell business users about effects that are
unstable across different samples.
This point is particularly relevant for systems
that combine data miners with optimizers that
make use of any stochastic component.

5. No data miner works best for all applications~\cite{Wolpert96}.
Hence, we offer the same advice as with 
point\#4,5 in \S\ref{sec:12tipso}. That is,
 when faced with a new problem, it is useful to try several data miners and
stop after two significant improvements have been achieved over some initial baseline result.
In terms of what data miners
 to try first,
there is a large candidate list. For
software analytics, we refer the reader to table IX of
Ghotra et al.~\cite{ghotra15} that ranks dozens of different
data mining algorithms into four ``ranks''.
To sample a wide range of algorithms,
we suggest applying one algorithm for each rank.

6. Watch out for the temporal effect of data -- it may cause past models to become inadequate.

7. Strive to avoid overfitting.
Try to test on data not used in training.
If the data has timestamps, train on earlier data and test on later data. If no timestamps, then ten times randomly reorganized the data and divide it into ten bins.
Next, make each bin the test set and all the other bins the train set. 

8. Ignore spurious
distinctions in the data. For example,  
the Fayyad-Irani discretizer~\cite{fayyad1993multi} can simplify numeric columns by dividing
up  regions that best divide up the target class.

9. Ignore spurious columns.
If a  column ended up being poorly discretized, that is a symptom that  that column
is uninformative. By pruning the
columns with low discretization scores, spurious data
can be ignored~\cite{hall03}.

10. Ignore spurious rows.
Similarly, instance and  range pruning can be useful. After discretization and feature selection, numeric ranges can be scored by how well they achieve specific target classes. 
 If we delete rows that have few interesting ranges, we can reduce and simplify any process that visualizes or searches the data~\cite{peters2015lace2}.

11. If there is very little data, consider asking the model inside the data miner to generate more examples. If that is not practical (e.g. the model is too slow 
to execute) then try transfer learning~\cite{Krishna18}, active learning~\cite{yu2018finding},
or semi-supervised learning.

12. For more advice about using data miners,
see ``Bad Smells for Software Analytics''`\cite{MENZIES201935}.

\BLACK

\section{Claim4:  {\RED For software engineering tasks,\BLACK}  data mining without optimization is not recommended}
\label{sec:claim5}\label{tion:claim5}
 

There are many reports in the empirical  SE literature
where the results of a data miner are used to defend claims such as: 
\bi
\item
{\em ``In this domain, the most important concepts are X.''}
For example, Barua et al.~\cite{Barua2012WhatAD} used  text mining   called  to conclude
what topics are most discussed at Stack Overflow.
\item
{\em ``In this domain, learnerX is better than learnerY
for building models.''} For example, Lessmann et al.~\cite{lessmann8} reported that the CART decision tree
performs much worse than random forests for
defect prediction.
\ei
All the above results were generated using
the default values  for CART, random forests and a particular text mining algorithm. We note that these conclusions are now questionable given that tuned learners produce very different results to untuned learners.
For example:
\bi
\item Claims like {\em ``In this domain, the most important concepts are X''} can be 
changed by applying an optimizer to a data miner. For example,
Tables 3 and 8  of~\cite{57agrawal2018wrong}
show what was found before/after tuned
text miners were applied to Stack Overflow data. In 
many cases, the pre-tuned topics just disappeared after tuning.  Also,
in defense of the tuned results, we note that, in ``order effects experiments'', 
the pre-tuned topics were far more ``unstable''
than the tuned topics\footnote{In ``order effects experiments'', the training data is  re-arranged at random  before running the learner again. In such experiments,
a result is ``unstable''
if the learned model changes just by re-ordering the training data.}.
\item Claims like {\em  ``In this domain, learnerX is better than learnerY
for building models''} can be changed  by tuning.
One example Fu et al. reversed some of the Lessmann et al. conclusions
by showing that  tuned CART  performs much better than random forest~\cite{35fu2016}.
For another example, recall \fig{tanit} where, before/after tuning the C5.0 algorithm was the
worst/best learner (respectively).
\ei

\begin{figure}
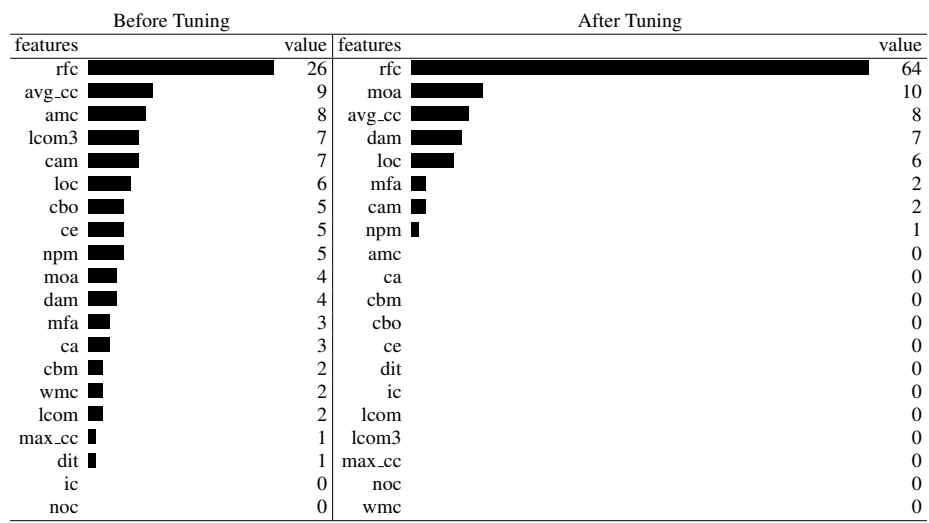

\begin{center}
\resizebox{\textwidth}{!}{
\begin{tabular}{rlr|rlr}
\multicolumn{3}{c}{Before Tuning} & \multicolumn{3}{c}{After Tuning} \\ \hline
 features & ~ & value & features & ~ & value \\\hline
rfc & \textcolor{black}{\rule{26mm}{2mm}}  &  26 & rfc & \textcolor{black}{\rule{64mm}{2mm}}  &  64 \\
avg\_cc & \textcolor{black}{\rule{9mm}{2mm}}  &  9 & moa & \textcolor{black}{\rule{10mm}{2mm}}  &  10 \\
amc & \textcolor{black}{\rule{8mm}{2mm}}  &  8 & avg\_cc & \textcolor{black}{\rule{8mm}{2mm}}  &  8 \\
lcom3 & \textcolor{black}{\rule{7mm}{2mm}}  &  7 & dam & \textcolor{black}{\rule{7mm}{2mm}}  &  7 \\
cam & \textcolor{black}{\rule{7mm}{2mm}}  &  7 & loc & \textcolor{black}{\rule{6mm}{2mm}}  &  6 \\
loc & \textcolor{black}{\rule{6mm}{2mm}}  &  6 & mfa & \textcolor{black}{\rule{2mm}{2mm}}  &  2 \\
cbo & \textcolor{black}{\rule{5mm}{2mm}}  &  5 & cam & \textcolor{black}{\rule{2mm}{2mm}}  &  2 \\
ce & \textcolor{black}{\rule{5mm}{2mm}}  &  5 & npm & \textcolor{black}{\rule{1mm}{2mm}}  &  1 \\
npm & \textcolor{black}{\rule{5mm}{2mm}}  &  5 & amc & \textcolor{black}{\rule{0mm}{2mm}}  & 0 \\
moa & \textcolor{black}{\rule{4mm}{2mm}}  &  4 & ca & \textcolor{black}{\rule{0mm}{2mm}}  & 0 \\
dam & \textcolor{black}{\rule{4mm}{2mm}}  &  4 & cbm & \textcolor{black}{\rule{0mm}{2mm}}  & 0 \\
mfa & \textcolor{black}{\rule{3mm}{2mm}}  &  3 & cbo & \textcolor{black}{\rule{0mm}{2mm}}  & 0 \\
ca & \textcolor{black}{\rule{3mm}{2mm}}  &  3 & ce & \textcolor{black}{\rule{0mm}{2mm}}  & 0 \\
cbm & \textcolor{black}{\rule{2mm}{2mm}}  &  2 & dit & \textcolor{black}{\rule{0mm}{2mm}}  & 0 \\
wmc & \textcolor{black}{\rule{2mm}{2mm}}  &  2 & ic & \textcolor{black}{\rule{0mm}{2mm}}  & 0 \\
lcom & \textcolor{black}{\rule{2mm}{2mm}}  &  2 & lcom & \textcolor{black}{\rule{0mm}{2mm}}  & 0 \\
max\_cc & \textcolor{black}{\rule{1mm}{2mm}}  &  1 & lcom3 & \textcolor{black}{\rule{0mm}{2mm}}  & 0 \\
dit & \textcolor{black}{\rule{1mm}{2mm}}  &  1 & max\_cc & \textcolor{black}{\rule{0mm}{2mm}}  & 0 \\
ic & \textcolor{black}{\rule{0mm}{2mm}}  & 0 & noc & \textcolor{black}{\rule{0mm}{2mm}}  & 0 \\
noc & \textcolor{black}{\rule{0mm}{2mm}}  & 0 & wmc & \textcolor{black}{\rule{0mm}{2mm}}  & 0 \\\hline
\end{tabular}
}
\end{center}
\caption{Features importance shown for decision tree before and after tuning on jEdit defect prediction where it is optimized for {\em minimizing} distance to ``utopia'' (where ``utopia''
is the point recall=1, false alarms=0).}
\label{fig:feature_importance}
\end{figure}

For a small example of this effect
(that optimizing a data miner leads to
different results), see
Figure~\ref{fig:feature_importance}. This figure
ranks the importance of different static code features in
a defect prediction decision tree.
Here ``importance'' is computed as the (normalized) total reduction of the Gini index for a feature\footnote{The Gini index measures class diversity after a set of examples is divided by some criteria -- in this case,
the values of an attribute.}.
In this case, tuning significantly improved the performance of the learner (by 16\%, measured in terms of the ``utopia'' index\footnote{The distance to of a predictor's performance to the   ``utopia''
  point of recall=1, false alarms=0.}).
After tuning:

\bi
\item
Features that seemed irrelevant   (before tuning) are found to be very important
(after tuning); e.g. moa, dam.
\item
Many features received much lower ranking (e.g. amc, lcom3, cbo).
\ei
Overall, the number of ``interesting'' features that we might choose to report as ``conclusions'' in this study is greatly reduced.

Apart from the examples in this section, there  are many other examples where (a)~hyperparameter optimization
 selects   models with better performance and (b)~those selected models report very different things to alternate models.
 For example, in \tion{fastop}, we saw the following: 
\bi
\item
In the   Nair et al.~\cite{87nair2017using} case study, the  CART decision tree was used to summarize
the data seen so far. Such decision trees usually prune away most variables so that any human reading a CART
model would see different things than if they read a logistic regression model (where every variable may appear in the logistic equation).
\item
In the
Tantithamthavorn et al.~\cite{47tantithamthavorn2016automated} case study of \fig{tanit}, depending on what learner
was selected by what optimizer, that analysis  would have reported models as a single decision tree, a forest of trees,
a set of rules, the probability distributions within a Bayes classifiers, or as an opaque neural net model.
\item
In the Majumder et al.~\cite{88majumder2018500+} case study,  tuning made us select
SVM over a deep learner.
Note that that change (from one learner to another) also changes what we would report from that model. 
 SVM models can be reported as the difference between their support vectors (which are 
few in number). A report of
the deep learning model may be much more complicated (since such a summary would  require   any number of complex transforms,   none of which are guaranteed to endorse the same model as the SVM),
\item In the Chen et al.~\cite{31chen2018,89chen2018sampling} case study, tuning used a recursive
clustering algorithm (that  prune away most of the details in the original model). Such pruning is a very different approach to 
that seen in other kinds of 
optimizers. Hence, any model learned from the Chen et al. methods would be very different to models learned from  other optimizers.
\ei
Note that the above effect, where the nature of the model generated by a learner is effected by the optimizer attached
to that learner, is quite general to all machine learning algorithms. 
F{\"u}rnkranz
and Flach characterize learners as ``surfing'' a landscape of modeling options looking for a ``sweet spot''
that balances different criteria (e.g. false alarms vs recall). Figure~2 offers an insightful
example of this process. In that figure, a learner adds more and more conditions to a model, thereby driving it to
different places on the landscape.
Depending on the hyperparameters of a learner, that learner will
``surf'' that space in different ways, terminate at different locations, and return different models.

The important point here is that, in domains where data miners can be optimized very quickly,
it would take just a few minutes to hours
to refine and improve untuned results. Further, when humans lean in to read what has been learned, these different
tunings lead to different kinds of models and hence different kinds of conclusions.
Hence, we do not  recommend reporting on the models learned via
 data mining without optimization.
 
It is worth noting that the use of optimizers to tune the learners' hyperparameters does not mean that we should ignore information on the range of results achieved using different hyperparameter choices. Analyses of sensitivity to hyperparameters are still important, especially considering that the best hyperparameters ``right now'' may not be the best hyperparameters for ``later''. Moreover, optimizers have themselves' hyperparameters, which could potentially affect their ability to suggest good hyperparameters to learners. Therefore, it is important to understand how sensitive the learners are to hyperparameter choice.
 \BLACK



 \section{Research Directions}

One way to assess any proposed framework such as {\IT} is as follows:
is it sufficient to guide the on-going work of a large community of researchers? As argued in this section, {\IT} scores very well on this criterion.

Firstly, given  that {\em \RED for software engineering tasks, data mining without optimization should be deprecated}, it is time to revisit and recheck every prior software analytics result based on untuned data miners. This will be a very large task that could consume the time of hundreds of graduate students for at least a decade.

Secondly, given that {\em \RED for software engineering tasks, data miners can greatly improve optimizers}, there are many research directions:
\bi
\item  {\em Better explanation and comprehension tools for AI systems:} Use the data miners to summarize complex optimizer output in order to convert opaque inference into something comprehensible. Some methods for that were seen above (\fig{veer} and the STAR algorithm of \S\ref{sect:betterco})
but they are just two early prototypes. Adding comprehension tools to AI systems that use optimization
is a fertile ground for much future research. For
a discussion on criteria to assess comprehensibility in software
analytics, see~\cite{Mori2018,Chen2018ApplicationsOP}.
\item {\em Auditable AI:} There is much recent interest in allowing humans to query AI systems for their biases and, where appropriate, to adjust them. Sampling tools like those of Chen and Nair et al.~\cite{31chen2018,89chen2018sampling,90nair2018finding}  offer  functionality that might be particularly suited for  that task. Recall that these tools optimized their models using just a few dozen examples -- which is a number small enough to enable human inspection. Perhaps we could build human-in-the-loop systems where humans watch the samplers' explored options -- in the field of optimization, the concepts of interactive optimization, dynamic optimization, and the multi-objective encoding of user preferences is well-established.
At the very least, this might allow humans to understand the reasoning.
And at the very most, these kinds of tools might allow humans to ``reach  in'' and better guide the reasoning.
\item {\em Faster Deep Learning:}
One open and pressing issue in software analytics is the tuning of deep learning networks. 
Right now, deep learning training is so slow that it is common practice to download pre-tuned networks. This means that  deep learning for software analytics may be prone to all the problems we discussed in claims 2 and 4. We gave some ideas here on how data mining can divide up and simplify the deep learning training problem (but, at this time, no definitive results).
\item
{\em Avoiding Hyper-hyperparameter Optimization:} While hyperparameter optimization is useful, it should be noted that the default parameters of the hyperparameter optimizers may themselves need optimizing by other optimizers. This is a problem since if hyperparameter optimization is slow, then hyper-hyperparameter optimization would be even slower. 
So how can we avoid  hyper-hyperparameter optimization?
\bi
\item
One possible approach is {\em transfer learning}. Typically when something is tuned, we do so because it will be run multiple times. So instead of search
{\em taula rasa} for good tunings, perhaps it is possible to partially combine parts of
old solutions to speed up the search for good hyperparameter values~\cite{Nair2018TransferLW}.   
\item
An analogous approach is to   select from a portfolio of algorithms. This typically involves the training of machine learning models on performance data of algorithms in combination with instances given as feature data. In software engineering, this has been recently used for the Software Project Scheduling Problem \cite{Shen2018,wu2016}. The field of per-instance configuration has received much attention recently, and we refer the interested reader to a recent updated survey article~\cite{kotthoff2016survey}. 
\item
Another approach is to find shortcuts around the  optimization process. For recent work on that, which we would characterize as highly speculative, see~\cite{fu2018building}.
\ei
\ei
Thirdly, given that  {\em \RED for software engineering tasks,{\BLACK} optimizers can greatly improve data miners}, it is time to apply hyperparameter optimization to more data mining tasks within software analytics. As shown by this paper, such an application can lead to dramatically better predictors.
Moreover, given that multi-objective perspective that can be given by optimizers, we can use optimizers to enable data mining to explore multiple objectives. For example:
\bi
\item
For software analytics, we could try to learn data miners that find the highest priority bugs 
after the {\em fewest} tests, found in the {\em smallest} methods in code that is {\em most familiar} to the current human inspector.   Such a data miner would return the most important bugs and easiest to fix (thus reducing issue resolution time for important issues).
\item
For project management,
when crowdsourcing large software projects, we could allocate tasks to programmers in order to {\em minimize} development time while {\em maximizing} work assignments to programmers that have the  most familiarity with that area of the code.
\item
For refereeing new research results in SE, the tools described here could  assign reviewers to new results in order to {\em minimize} the number of reviews per reviewer while {\em maximizing} the number of reviewers who work in the domain of that paper.
\ei

\noindent We could also extend SE to make it about using DUO for providing software engineering support for artificial intelligence
systems. For example, using the optimization discussed above, Weise et al.~\cite{Weise2018AnIG} have run over 157 million experiments on over 200 instances of two classical AI combinatorial problems. They found that the naive configuration is a good baseline approach, but with effort, it was possible to outperform it. 
Friedrich et al.~\cite{Friedrich2018EscapingLarge,Friedrich2019heavytailed} studied a
particular family of heuristic hill-climbers problems.
Their empirical optimization results sparked extensive theoretical investigations (i.e., proper computational complexity analyses) that showed that the new algorithm configuration is provably faster than what has been state-of-the-art. These can be seen as examples of software engineering to find better configurations. Other work in this area includes
Neshat et al.~\cite{Neshat2018waveenergy} who studied wave energy equations. 
Their optimization results were translated into well-performing algorithms for large problems. This can be seen as software engineering to improve algorithms' performance.

Lastly, given  {\em that \RED for software engineering tasks,{\BLACK} data mining and optimization is
essentially the same thing}, it is time to explore engineering principles for creating unified data mining/optimizer toolkits. We already have one exemplar of such a next-generation toolkit: see \url{www.automl.org} for the work on AutoML and its connections to Weka and scikit-learn. That said, this research area  is wide open for experimentation. For two reasons, we would encourage researchers to ``roll their own''
{\IT} implementations before automatically turning to tools like AutoML:
\bi
\item Some initial results suggest it may not be enough to just use AutoML~\cite{Tu2018IsOH} (in summary, given $N$ hyperparameter optimizers, AutoML was ``best'' for a minority of goals and datasets). 
\item In terms of training ourselves on how easy it is to combine optimizers and data miners, there is no substitute for ``rolling your own'' implementation (at least once, then perhaps moving on to tools like AutoML).
\ei

\RED
 \section{Related Work}\label{shoon}
 
 As discussed in \S\ref{sec:claim1}, this paper is a  reflection of two related research
 areas which, currently, are explored separately by different research teams.
 It is not the only attempt to do that. 
At the 2017 NII Shonan Meeting on ``Data-Driven Search-Based Software Engineering''~\cite{wagner17sh},
 three dozen researchers from the software analytics and search-based software engineering communities have met to revisit the 2003 technical note 
 ``Reformulating software engineering as a search problem'' by Clarke et al.~\cite{clarke2003reformulating}.
 While some of the attendees discussed the mechanics
of search-based methods and how they can help software analytics, most of the discussions focused on how to exploit all the data about software projects that has recently become available (e.g. at online sites like Github, etc) to help search-based software engineering. Such data could form valuable priors. For example:
 \bi
 \item
 Researchers exploring cross-project defect prediction report that lessons learned
 from one project can now be applied to another~\cite{zhang2016cross, ryu2016value, peters2015lace2, kamei2016studying, krishna2018bellwethers}.
 \item
 Researchers exploring language models in SE reported highly repetitive regularities
 in source code, making that artifact most amenable to prediction of nominal and 
 off-nominal behaviours~\cite{jensen2019naturalness,hellendoorn2018naturalness,allamanis2018survey}.
 \ei
 This paper differs from the above in that we explore the advantages of new algorithms (optimizers) for software analytics, while much of the discussions at Shonan were about the advantages of new data.
Given that, a natural future direction would be to combine both the new algorithms discussed here with the new
data sources. 
 \BLACK
\section{Conclusion}

For software analytics it is 
{\em possible}, {\em useful} and 
{\em recommended} to combine data mining and optimization using {\IT}. Such combination can lead to better (e.g., faster, more accurate or reliable, more interpretable, and multi-goals) analyses in empirical software engineering studies, in particular those studies that require automated tools for analyzing (large quantities of) data. We support our arguments based on a literature review of applications of {\IT}.

We say it is {\em possible} to combine data mining and optimization since data mining and optimization perform essentially the same task (Section \ref{sec:claim1}). Hence, it is hardly surprising that it can be a relatively simple matter to build a unified data mining/optimizer tool. For instance:
\bi
\item Nair et al.'s approach~\cite{90nair2018finding} was just a for-loop around CART.
\item STAR~\cite{Menzies:2007} was also a very simple learner (see \S\ref{sect:betterco}) wrapped around a simulated annealer, then re-ran the simulated annealer after setting the first $i$-best ranges.
\item Sampling  is just a simple bi-recursive clustering algorithm (but see {\S}3.4 of~\cite{89chen2018sampling} for a discussion on some of the nuances of that process).
\item There are many mature open source data mining and optimization frameworks\footnote{E.g. in Python: scikit-learn and DEAP~\cite{scikit-learn,rainville2012deap}.
E.g. in Java: Weka and (jMetal or SMAC)~\cite{Hall:2009,Durillo2011,hutter2011sequential}.
}. Also, some of the data mining and optimization algorithms 
are inherently very simple to implement (e.g. naive Bayes~\cite{Frank2000}, differential evolution~\cite{Storn:1997}). Hence it is easy to implement  optimizer+data miner combinations.
\ei 
As to {\em usefulness}, this paper has listed several benefits of {\IT}:
\bi
\item Data miners can speed up optimizers by dividing large problems into several  simpler and smaller ones. 
\item Also, when optimizers return many example solutions, data miners can generalize and summarize those into a very small set of exemplars (see for example \fig{veer}) or rules (see for example the STAR algorithm of \S\ref{sect:betterco}).
\item Optimization technology lets data miners explore a much broader set of competing goals than just (e.g.) precision, recall, and false alarms. Using those goals, it is possible
to better explore an interesting range of SE problems such as 
project management, requirements engineering, design, security problems,
software quality, software configuration,  mining textual SE artifacts,
just to name a few. 
\ei
Finally, as to {\em recommended}, we warn that it can be misleading to report
conclusions from an untuned learner since those conclusions can changed by tuning.
Since papers that use untuned learners can be so easily refuted, this community should be wary of publishing analytics papers that lack an optimization component.

Having made this case, we acknowledge that {\IT} would require  a paradigm shift for the software analytics community. Graduate subjects would have to be changed (to focus on different kinds of algorithms and case studies); toolkits would need to be reorganized; and journals and  conferences will have to adjust their paper selection criteria.
Looking into the future, we anticipate several years where {\IT} is explored
by a minority of software analytics researchers. That said, by 2025,
 we predict that {\IT} will be standard  practice in software analytics.

\section*{Acknowledgements}

Earlier work ultimately leading to the present one was inspired by the NII Shonan Meeting on Data-Driven Search-based Software Engineering (goo.gl/f8D3EC), December 11-14, 2017.
We thank the organizers of that workshop (Markus Wagner, Leandro L. Minku, Ahmed E. Hassan, and John Clark) for their academic leadership and inspiration. 

Dr Menzies' work was partially supported by NSF grant No. 1703487.

Dr Minku's work was partially supported by EPSRC grant Nos. EP/R006660/1 and EP/R006660/2.

Dr Wagner's work was partially supported by the ARC grant DE160100850.

\bibliographystyle{spmpsci}

\section*{}

\begin{wrapfigure}{l}{25mm} 
    \includegraphics[width=1in,clip,keepaspectratio]{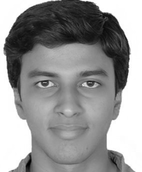}
  \end{wrapfigure}\par\textbf{Amritanshu Agrawal} holds a Ph.D. in Computer Science from North Carolina State University, Raleigh, NC.
He explored better and faster hyperparameter optimizers for software analytics. He works as a Data Scientist at Wayfair, Boston. For more, see \url{http://www.amritanshu.us}.
\\
\\
\\
\\
\\
\section*{}

\begin{wrapfigure}{l}{25mm} 
    \includegraphics[width=1in,clip,keepaspectratio]{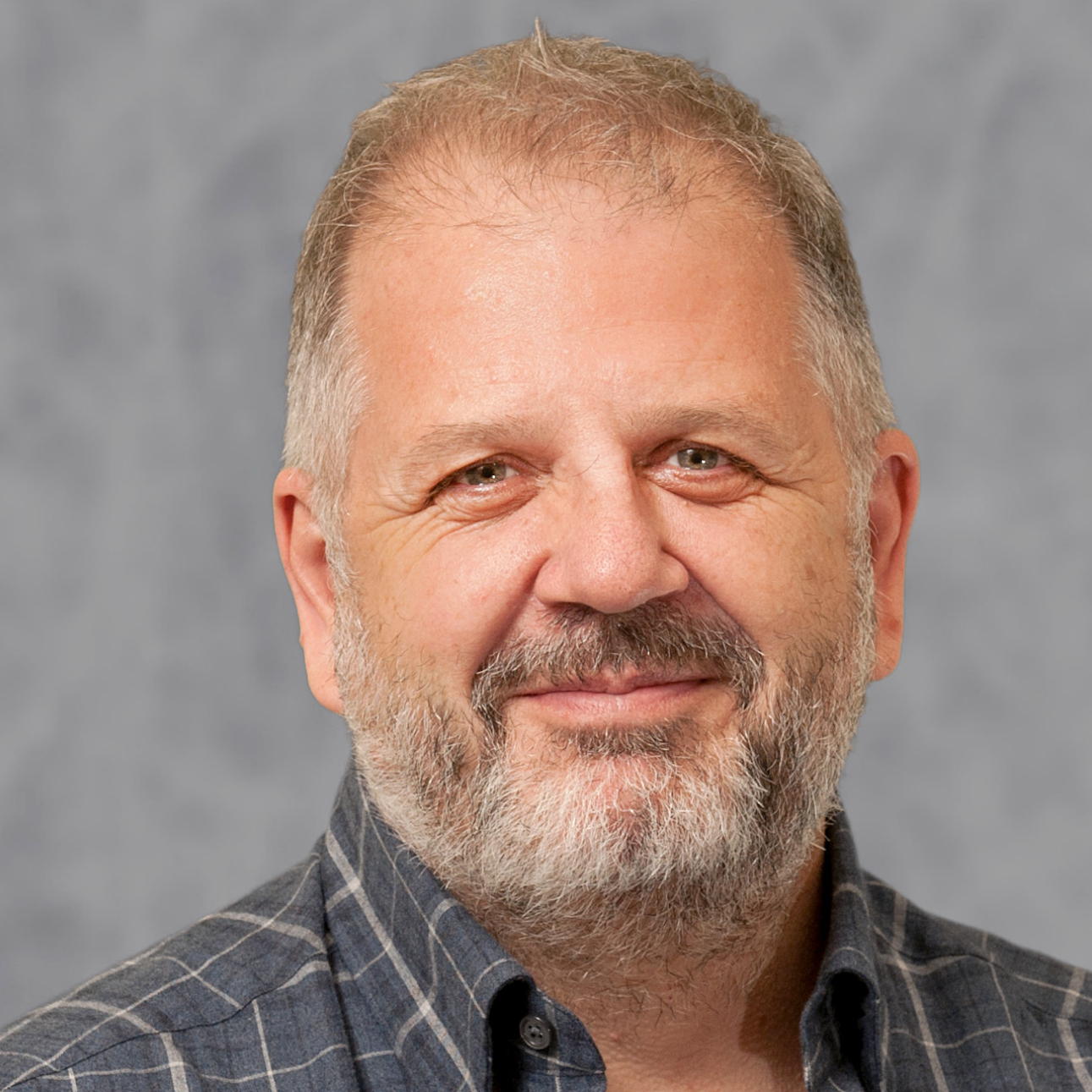}
  \end{wrapfigure}\par
  \textbf{Dr. Tim Menzies} is a Professor in CS at NcState.  His research interests include software engineering (SE), data mining, artificial intelligence, search-based SE, and open access science. \url{http://menzies.us}
\\
\\
\\
\\
\\
\section*{}

\begin{wrapfigure}{l}{25mm} 
    \includegraphics[width=1in,clip,keepaspectratio]{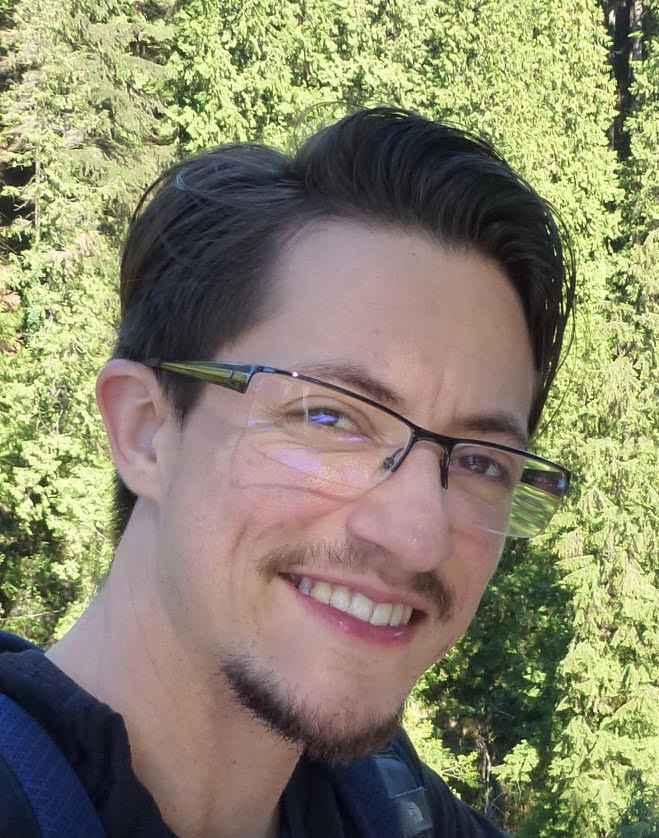}
  \end{wrapfigure}\par
  \textbf{Dr. Leandro L. Minku} is a Lecturer in Intelligent Systems at the School of Computer Science, University of Birmingham (UK). Prior to that, he was a Lecturer in Computer Science at the University of Leicester (UK), and a Research Fellow at the University of Birmingham (UK). He received the PhD degree in Computer Science from the University of Birmingham (UK) in 2010. Dr. Minku's main research interests are machine learning for software engineering, search-based software engineering, machine learning for non-stationary environments / data stream mining, and ensembles of learning machines. His work has been published in internationally renowned journals such as IEEE Transactions on Software Engineering, ACM Transactions on Software Engineering and Methodology, IEEE Transactions on Knowledge and Data Engineering, and IEEE Transactions on Neural Networks and Learning Systems. Among other roles, Dr. Minku is general chair for the International Conference on Predictive Models and Data Analytics in Software Engineering (PROMISE 2019 and 2020), co-chair for the Artifacts Evaluation Track of the International Conference on Software Engineering (ICSE 2020), and associate editor for IEEE Transactions on Neural Networks and Learning Systems, Journal of Systems and Software, and Neurocomputing journals.
\section*{}

\begin{wrapfigure}{l}{25mm} 
    \includegraphics[width=1in,clip,keepaspectratio]{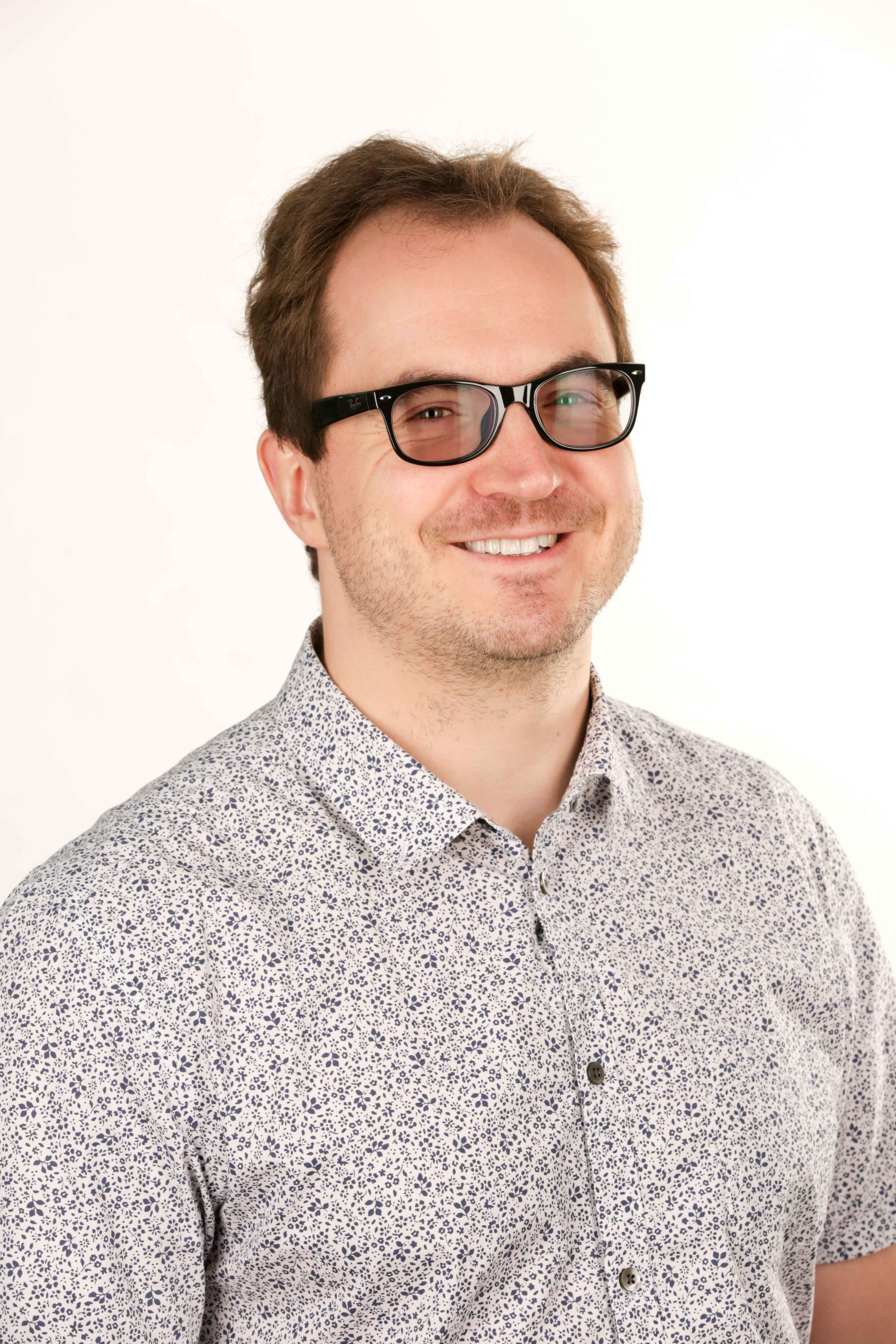}
  \end{wrapfigure}\par
  \textbf{Dr. Markus Wagner} is a Senior Lecturer in optimisation and logistics at the School of Computer Science, University of Adelaide, Australia. He is interested in the broad topic of heuristic optimisation, which for him ranges from theoretical investigations and theory-motivated algorithm design to applications in software engineering, such as, non-functional code optimisation and artefact summarisation. He has been the lead organiser of the NII Shonan Meeting on Data-Driven Search-based Software Engineering (goo.gl/f8D3EC, December 11-14, 2017), which has ultimately resulted in this very article. Two fun facts: he has over 120 co-authors, and he is ACM SIGEVO’s first Sustainability Officer. \url{https://cs.adelaide.edu.au/~markus/}
  \\
\section*{}

\begin{wrapfigure}{l}{25mm} 
    \includegraphics[width=1in,clip,keepaspectratio]{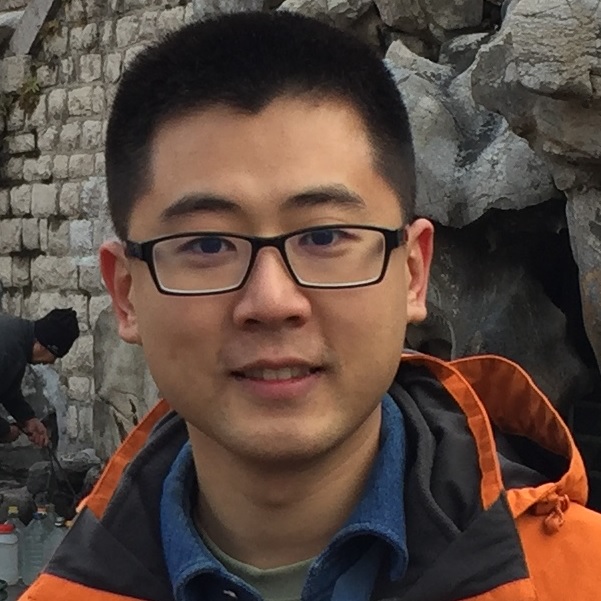}
  \end{wrapfigure}\par
  \textbf{Zhe Yu}
 is a fifth year Ph.D. student in Computer Science at North Carolina State University. He received his masters degree in Shanghai Jiao Tong University, China. His primary interest lies in the collaboration of human and machine learning algorithms that leads to better performance and higher efficiency. \url{http://azhe825.github.io/}


\end{document}